\documentclass[12pt]{article}
\begin{document}
\newfont{\elevenmib}{cmmib10 scaled\magstep1}%
\renewcommand{\theequation}{\arabic{section}.\arabic{equation}}

\newcommand{\preprint}{
            \begin{flushleft}
   \elevenmib Yukawa\, Institute\, Kyoto\\
            \end{flushleft}\vspace{-1.3cm}
            \begin{flushright}\normalsize  \sf
            YITP-99-19\\
            DTP/99/31\\
   {\tt hep-th/9905011} \\ May 1999
            \end{flushright}}
\newcommand{\Title}[1]{{\baselineskip=26pt \begin{center}
            \Large   \bf #1 \\ \ \\ \end{center}}}
\newcommand{\Author}{\begin{center}\large \bf
            A.\, J.\, Bordner$^a$, E.\, Corrigan$^b$, 
	    and R.\, Sasaki$^a$ \end{center}}
\newcommand{\Address}{\begin{center} \it
            $^a$ Yukawa Institute for Theoretical Physics, Kyoto
            University,\\ Kyoto 606-8502, Japan \\
	    $^b$ Department of Mathematical Sciences, University of
            Durham,\\ South Road, Durham DH1-3LE, United Kingdom
      \end{center}}
\newcommand{\Accepted}[1]{\begin{center}{\large \sf #1}\\
            \vspace{1mm}{\small \sf Accepted for Publication}
            \end{center}}
\baselineskip=20pt

\preprint
\thispagestyle{empty}
\bigskip
\bigskip
\bigskip
\Title{Generalised Calogero-Moser models and \\
 universal Lax pair operators}
\Author

\Address
\vspace{2cm}

\begin{abstract}
Calogero-Moser models can be generalised for all of the finite
reflection  groups.
These include models based on non-crystallographic root systems, that is
the root systems of the finite reflection groups,  $H_{3}$,
 $H_{4}$, and the dihedral group $I_{2}(m)$, besides the well-known
ones based on crystallographic root systems, namely those associated with
Lie algebras.
Universal Lax pair operators for all of the generalised Calogero-Moser models
and for any choices of the potentials are constructed as linear
combinations of the reflection operators.
The consistency conditions are reduced to functional equations for the
coefficient functions of the reflection operators in the Lax pair.
There are only four types of such functional equations corresponding to
the two-dimensional sub-root systems, \(A_2\),  \(B_2\), \(G_2\), and 
\(I_2(m)\). The root type and the minimal type Lax pairs, derived in our
previous papers, are given as the simplest representations.
The spectral parameter dependence plays an important role in the Lax
pair operators, which bear a strong resemblance to the Dunkl operators, a
powerful tool for solving quantum Calogero-Moser models.

\end{abstract}

\section{Introduction}
\label{intro}
\setcounter{equation}{0}
Generalised Calogero-Moser models are integrable many-particle 
dynamical systems
based on finite reflection groups. 
Finite reflection groups include
 the dihedral groups \(I_2(m)\) and \(H_3\) and \(H_4\)
together with the Weyl groups of the root systems associated with Lie
algebras, called crystallographic root systems.
Integrability of classical Calogero-Moser models based on the
crystallographic root systems \cite{CalMo,OP1} is shown in terms
of Lax pairs.
The root and the minimal type Lax pairs derived in our previous papers 
\cite{Our_CM_Papers} provide a universal framework for these
Calogero-Moser models, including those based on exceptional root systems
and the twisted models.
On the other hand, a theory of classical integrability for the models
based on non-crystallographic root systems has been virtually
non-existent.
This is in sharp contrast with the quantum counterpart.
Dunkl operators, which are useful for solving quantum Calogero-Moser
models, were first explicitly constructed for the models based on the
dihedral groups \cite{Dunk}.

In this paper we present a Lax pair in an operator form for generalised
Calogero-Moser models, which applies universally to the models based on
non-crystallographic root systems as well as those based on
crystallographic ones.
This Lax pair, as expected, bears a strong resemblance to the Dunkl
operators \cite{Dunk,BFV} and the reflection operators play a central
role.
The spectral parameter dependence is also essential, in contradistinction
to the auxiliary role in the conventional formalism \cite{OP1}.
When suitable representation spaces are chosen, the universal Lax pair 
reproduces the root type and the minimal type Lax pairs for
the models based on the crystallographic root systems.
This provides another clue that the quantum and classical
integrability of the generalised Calogero-Moser models are  closely
connected.
We hope that this Lax pair operator formalism will cross-fertilise the
fruitful subject of the quantum and classical Calogero-Moser systems and
related ones such as Toda systems.

For the general background and the motivations of this
 paper and the physical applications of the
Calogero-Moser models with various potentials to
lower-dimensional physics, ranging from solid state to
particle physics and   supersymmetric gauge theories, we refer to our
previous papers \cite{Our_CM_Papers} and references therein.

This paper is organised as follows. In section \ref{rootsystems}
we summarise the basic concepts of the finite reflection groups in order to
set the stage and to introduce appropriate notation.
In section \ref{gencal-mo} the generalised Calogero-Moser models are
defined with various choices of  root systems and  potentials.
Section \ref{consistency} is the main body of the paper. The Lax operators
\(L\) and \(M\) are defined as a linear combination of the reflection
operators \(\hat{s}_{\rho}\) for all the roots \(\rho\).
The coefficient functions depend on the dynamical variables and on the
spectral parameter in a rather symmetrical way.
Consistency of the Lax pair can be proved quite easily.
The Lax equation is at most quadratic in the reflection operators,
 \(\hat{s}_{\rho}\hat{s}_{\sigma}\), which becomes an identity operator
for 
\(\rho=\sigma\), \(\hat{s}_{\rho}^2=1\), and a two-dimensional rotation operator
for \(\rho\neq\sigma\).
The linear (\(\hat{s}_{\rho}^1\)) and the constant
(\(\hat{s}_{\rho}^0\)) parts
give the canonical equations of motion of the generalised Calogero-Moser
models. The quadratic part \(\hat{s}_{\rho}\hat{s}_{\sigma}, \rho\neq\sigma\),
imposes the consistency conditions, which are decomposed into those
corresponding to two-dimensional sub-root systems containing 
\(\rho\) and \(\sigma\).  As shown in Table 1, 
 there are only
four types of two-dimensional root systems, \(A_2\),  \(B_2\), \(G_2\),
and 
\(I_2(5)\) for all of the Coxeter groups except for the dihedral group
\(I_2(m)\) which can have many for some values of \(m\). 
Thus the functions appearing in the Lax pair operator (except for those
for \(I_2(m)\)) need to satisfy at most two functional equations.
The solutions are derived in the Appendix.
In section \ref{reps} various possible representations of the Lax pair
operator are discussed.
The minimal type and the  root type Lax pairs are derived as two simplest
examples in section \ref{minroot}. Various sum rules utilised in previous
papers are derived as restrictions of the general functional equations
derived in section \ref{consistency}.
Some comments and discussion are given in section \ref{comdis}.
Details of the derivation of the solutions are relegated to the Appendix.
Two types of solutions, the untwisted and twisted solutions, are
derived.
The consistency conditions of all of the untwisted solutions are
attributed to one functional identity,  (\ref{Sigma_Fn_Identity}),
 of the Weierstrass \(\sigma\) function.
Twisted solutions are obtained as proper linear combinations of the
untwisted ones.

\section{Root systems and finite reflection groups}
\label{rootsystems}
\setcounter{equation}{0}

We now review some facts about root systems and their reflection
groups in order to introduce notation  \cite{Coxeter_groups}.  We
consider only
reflections in Euclidean space.
A root system $\Delta$ of rank \(r\) is a set of
vectors in $\mathbf{R}^{r}$ which is invariant under reflections 
in the hyperplane perpendicular to each
vector in $\Delta$.  In other words,
\begin{equation}
   s_{\alpha}(\beta)\in\Delta,\quad\forall \alpha,\beta\in\Delta, 
\end{equation}
in which 
\begin{equation}
   s_{\alpha}(\beta)=\beta-2(\alpha\cdot\beta/|\alpha|^{2})\alpha.
\end{equation}
Dual roots are defined by $\alpha^{\vee}=2\alpha/|\alpha|^{2}$, in
terms of which
\begin{equation}
   \label{Root_reflection}
   s_{\alpha}(\beta)=\beta-(\alpha^{\vee}\!\!\cdot\beta)\alpha.  
\end{equation}
The orbit of $\beta\in\Delta$ is the set of root vectors
resulting from the action of the reflections on it
$\{s_{\alpha}(\beta), \alpha\in\Delta\}$.
The set of positive roots $\Delta_{+}$ may be defined in terms of a
vector $V\in\mathbf{R}^{r}$, with 
$V\cdot\alpha\neq 0,\,\forall\alpha\in\Delta$, as 
those roots $\alpha\in\Delta$ such that $\alpha\cdot V>0$.  A unique
set of $r$ simple roots $\Pi$ is defined such that they span
the root space and the coefficients $\{a_{j}\}$ in 
$\beta=\sum_{j=1}^{r}a_{j}\alpha_{j}$ for 
$\beta\in\Delta_{+},\,\{\alpha_{j}\in\Pi,\,j=1,\ldots,
r\}$ are all positive.

The set of reflections $\{s_{\alpha},\,\alpha\in\Delta\}$ generates a
group, known as a Coxeter group.  It is generated by products of
$s_{\alpha}$ with $\alpha\in \Pi$ subject to the relations
\begin{equation}
   \label{Coxeter_relations}
   (s_{\alpha}s_{\beta})^{m(\alpha,\beta)}=1,\qquad \alpha,\beta\in \Pi.
\end{equation}
The set of positive integers $m(\alpha,\beta)$ uniquely specify the 
Coxeter group with $m(\alpha,\alpha)=1,\,\forall
\alpha\in \Pi$. 
For example, for two-dimensional crystallographic root systems
\(A_2\), \(B_2\), and \(G_2\), the integer \(m(\alpha,\beta)\) is 
3, 4, and 6, respectively. Thus \(s_{\alpha}s_{\beta}\) is a 
two-dimensional rotation by an angle \(\pm{2\pi/3}\), \(\pm \pi/2\), 
and \(\pm\pi/3\), respectively. This fact will be used in later sections.
We consider here only those Coxeter groups with a finite 
number of roots in
Euclidean space, called the finite reflection groups.   

The root systems for finite reflection groups may be divided into two
types: crystallographic and non-crystallographic root systems.  
Crystallographic root systems satisfy the additional condition
\begin{equation}
   \alpha^{\vee}\!\!\cdot\beta\in\mathbf{Z},\quad \forall
   \alpha,\beta\in\Delta.
\end{equation}
These root systems are associated with simple Lie
algebras: \{$A_{r},\,r\ge 1\}$, $\{B_{r},\,r\ge 2\}$, $\{C_{r},\,r\ge 2\}$,
$\{D_{r},\,r\ge 4\}$, $E_{6}$, $E_{7}$, $E_{8}$, $F_{4}$, and
$G_{2}$ and  $\{BC_{r},\,r\ge 2\}$.  The Coxeter groups for these root
systems are called Weyl groups.  The remaining non-crystallographic root
systems are $H_{3}$, $H_{4}$, and the dihedral group of order $2m$, 
$\{I_{2}(m),\,m\ge 4\}$.  

Weyl chambers are defined as the open subsets of $\mathbf{R}^{r}$ that 
result from removing the reflection 
hyperplanes $H_{\alpha}$, $\alpha\in\Delta$, 
$H_{\alpha}\equiv\{q\in\mathbf{R}^{r},\,\alpha^{\vee}\!\!\cdot
q=0\}$.
The action of the reflection group on the Weyl
chambers is transitive and free, {\it i.e.} any Weyl chamber may be
obtained from another by the action of an element of the reflection
group and this element is unique.  The principal Weyl chamber is 
defined as the one whose points
$q$ satisfy $q\cdot\alpha>0,\,\forall\alpha\in\Delta_{+}$.
For crystallographic root systems this implies that all points in the
principal Weyl chamber have positive Dynkin labels.

Definitions and properties of the crystallographic root systems 
may be found in many references, see for example
\cite{Lie_Algebra_Ref}.  
The concept of weights may be defined for these root systems.
For crystallographic root systems, any positive root is a 
sum of simple roots in $\Pi$ 
with positive
integer coefficients.  
The set of weights $\Sigma$, which lie on a lattice, is the set of vectors
such that if $\Lambda\in \Sigma$ then $\alpha^{\vee}\!\!\cdot\Lambda$ 
is an integer for any 
$\alpha\in\Delta$.
Fundamental weights $\Lambda^{(j)}$ are 
vectors which form a dual basis to the corresponding dual simple roots 
$\alpha^{\vee}_{j}$, {\it i.e.}
$\alpha^{\vee}_{j}\cdot\Lambda^{(k)}=\delta_{jk},\,j,k=1,\ldots,r$.  Any
weight $\Psi\in \Sigma$ may be expressed as 
a sum of fundamental weights with integer
coefficients, $\Psi=\sum_{j=1}^{r}a_{j}\Lambda^{(j)}$.  The 
coefficients $\{a_{j},\,j=1,\ldots, r\}$ are called the Dynkin labels of
$\Psi$.

We now
briefly describe the non-crystallographic root systems. 
 The
dihedral group of order $2m$, $I_{2}(m)$, is the group of orthogonal
transformations that preserve a regular $m$-sided polygon in two
dimensions. It consists of $m$ rotations (through multiples of
$2\pi/m$) and $m$ reflections. The angle
between adjacent roots is
$\pi/m$ and a possible basis for the roots, if all are chosen to have the
same length $|\alpha_{j}|^{2}=1$, is
\begin{equation}
   \label{DihedralBasis}
   \alpha_{j}=\left(\cos(j\pi/m),\sin(j\pi/m)\right),\quad j=1,\ldots,2m.
\end{equation}
For odd $m$  all of the roots are in the same orbit of the reflection
group but for even $m$  there are two orbits, one consisting of the
$\alpha_{j}$ with odd $j$  and the other with even $j$ .  This then
allows two different coupling constants and potential functions for the
$I_{2}(m)$ Calogero-Moser model for even $m$.

The reflection group of type $H_{4}$ is the symmetry group of a
regular 120-side solid, with dodecahedral faces, in $\mathbf{R}^{4}$.
It is a group of order 14400.  
The group of type $H_{3}$ is a subgroup of $H_{4}$ and is 
the symmetry group of the icosahedron (with 20 faces) in
$\mathbf{R}^{3}$. It is a group of order 120. Define 
\begin{equation}
   a\equiv\cos{\pi\over 5}={1+\sqrt{5}\over 4},\qquad 
   b\equiv\cos{2\pi\over 5}={-1+\sqrt{5}\over 4}.
\end{equation}
Then a choice of simple roots for $H_{4}$ is the following:
\begin{eqnarray}
   \label{H4SimpleRoots}
   \alpha_{1}=\left(a,-{1\over 2},b,0\right), \qquad 
   \alpha_{2}=\left(-a,{1\over 2},b,0\right), \\
   \nonumber
   \alpha_{3}=\left({1\over 2},b,-a,0\right), \qquad 
   \alpha_{4}=\left(-{1\over 2},-a,0,b\right).
\end{eqnarray}
The full set of roots of $H_{4}$ in this basis may be obtained from
$(1,0,0,0)$, $({1\over 2},{1\over 2},{1\over 2},{1\over 2})$, 
and $(a,{1\over 2},b,0)$ by
even permutations and arbitrary sign changes of coordinates. 
These 120 roots form a single orbit.
A subset of (\ref{H4SimpleRoots}), $\{\alpha_{1},\alpha_{2},\alpha_{3}\}$
is a choice of simple roots for the $H_{3}$ root
system.  In this basis,  the full set of roots for $H_{3}$ results from
even permutations and arbitrary sign changes of $(1,0,0)$ and
$(a,{1\over 2},b)$. These 30 roots also form  a single orbit.

\section{Generalised Calogero-Moser Models}
\label{gencal-mo}
\setcounter{equation}{0}

A generalised Calogero-Moser model is a  
Hamiltonian system associated with a root system $\Delta$
of rank \(r\).  
Quantum versions of these models are also integrable, at least
for certain choices of $\Delta$ and potential function \cite{Quantum_CM}.
The dynamical variables are the coordinates
$\{q^{j}\}$ and their canonically conjugate momenta $\{p_{j}\}$, with
the Poisson brackets
\begin{equation}
  \{q^{j},p_{k}\}=\delta^{j}_{k},\qquad \{q^{j},q_{k}\}=
\{p_{j},p_{k}\}=0,\quad j,k=1,\ldots,r.
\end{equation}
These will be denoted by vectors in $\mathbf{R}^{r}$
\begin{equation}
  q=(q^{1},\ldots,q^{r}),\qquad p=(p_{1},\ldots,p_{r}).
\end{equation}
The Hamiltonian for the Calogero-Moser model is
\begin{equation}
   \label{CMHamiltonian}
   \mathcal{H} = {1\over 2} p^{2} + \sum_{\alpha\in\Delta} 
   {g_{|\alpha|}^{2}\over |\alpha|^{2}}
   \,V_{|\alpha|}(\alpha\cdot q),
\end{equation}
in which the real coupling constants $g_{|\alpha|}$ and potential
functions 
$V_{|\alpha|}$ are defined on orbits of the corresponding
finite reflection group, {\it i.e.} they are
identical for roots in the same orbit.  This then ensures that
the Hamiltonian is invariant under reflections of the phase space 
variables about a hyperplane perpendicular to any root
\begin{equation}
   q\rightarrow s_{\alpha}(q), \quad p\rightarrow s_{\alpha}(p), \quad
   \forall\alpha\in\Delta
\end{equation}
with $s_{\alpha}$ defined by (\ref{Root_reflection}).

The Lax pair operator that  we will construct in later sections will apply
for the following potentials (\(g_{|\alpha|}=g\) for all roots in simply 
laced models and  \(g_{|\alpha|}=g_L\) 
for long roots and \(g_{|\alpha|}=g_S\) for
short roots in 
non-simply laced models):
\begin{enumerate}
\item {\em Untwisted elliptic potential\/}. This applies to all of the
crystallographic root systems and the potential function is
\begin{equation}
   V_{|\alpha|}(\alpha\cdot q)=\wp(\alpha\cdot q|\{2\omega_1,2\omega_3\}),
   \quad \mbox{for all roots},
   \label{simppot}
\end{equation}
in which \(\wp\) is the Weierstrass \(\wp\) function with a pair 
of primitive periods
\(\{2\omega_1,2\omega_3\}\). 
Throughout this paper we adopt the convention that
the Weierstrass \(\wp\), \(\zeta\), and \(\sigma\) functions have the above
standard periods, unless otherwise stated.
\item {\em Twisted elliptic potential\/}. This applies to all of the
non-simply laced root systems. Except for the \(G_2\) model,
the potential functions are
\begin{equation}
   V_{|\alpha|}(\alpha\cdot q)=
   \cases{\wp(\alpha\cdot q|\{2\omega_1,2\omega_3\}),
   \quad \mbox{for long roots},\cr\cr
   \wp(\alpha\cdot q|\{\omega_1,2\omega_3\}),
   \quad\ \mbox{for short roots}.\cr}
\end{equation}
That is, the potential for short roots has one half of the standard period
in one direction, which we choose to be \(\omega_1\). For the \(G_2\)
model,
\begin{equation}
   V_{|\alpha|}(\alpha\cdot q)=
   \cases{\wp(\alpha\cdot q|\{2\omega_1,2\omega_3\}),
   \quad \mbox{for long roots},\cr\cr
   \wp(\alpha\cdot q|\{{2\omega_1\over3},2\omega_3\}),
   \quad\ \mbox{for short roots}.\cr}
\end{equation}
In this case the potential for short roots has one third of the standard 
period in one direction, which we choose to be \(\omega_1\). The cases of
\(BC_r\) system and the extended \(B_r\), \(C_r\), and \(BC_r\) systems
will be discussed separately in later sections.
\item
{\em Trigonometric and hyperbolic potentials\/}. This applies to all
crystallographic systems and the potential functions are
\begin{equation}
   V_{|\alpha|}(\alpha\cdot q)=
   \cases{{a^2/{\sin^2 a(\alpha\cdot q)}},
   \cr\cr {a^2/{\sinh^2 a(\alpha\cdot q)}},
   \cr}\quad \mbox{for all roots},
\end{equation}
in which \(a\) is an arbitrary real constant.
\item
{\em Rational potential\/}. This applies to all of the generalised
Calogero-Moser models including those based on the dihedral group
\(I_2(m)\), \(H_3\), and \(H_4\) and the potential function is
\begin{equation}
   V_{|\alpha|}(\alpha\cdot q)={1\over{(\alpha\cdot q)^2}},
   \quad \mbox{for all roots}.
   \label{ratpot}
\end{equation}
These models are also integrable if a confining harmonic potential
\begin{equation}
   {1\over2}\omega^2q^2
   \label{harmpot}
\end{equation}
is added to the Hamiltonian. 
\end{enumerate}

Some remarks are in order. (i) For all of the root systems and for 
any choice of the
potential,
the generalised Calogero-Moser model has a hard repulsive potential \(\sim
{1\over{(\alpha\cdot q)^2}}\) near the reflection hyperplane
\(H_{\alpha}=\{q\in\mathbf{R}^{r},\, \alpha^{\vee}\!\!\cdot q=0\}\).
This repulsion potential is classically insurmountable. 
Thus the motion is always 
confined within one Weyl chamber. 
In other words, the spatial ordering of the
particles is unchanged during the time evolution.
This simplifying feature is a basic cornerstone of the solvability.
The coupling constants \(g_{|\alpha|}^2\) 
(with a scale \({1/{|\alpha|^2}}\))
are  measures of the strength of the repulsive potentials. (ii) The
trigonometric, hyperbolic, and the rational potentials are obtained from
the elliptic potential as one or both periods tend to infinity.
The Lax pairs for these degenerate potentials can be obtained 
from the one for the
elliptic potential by taking the corresponding limit.
Thus we do not write down the Lax pairs for degenerate potential 
cases except for
the models based on the non-crystallographic root systems, 
for which only the
rational potentials are integrable. (iii) For all of the Lax pairs 
based on any root
systems and any choice of the potential, 
except for the rational potential with
the harmonic confining potential (\ref{harmpot}), one can introduce an
additional complex parameter  $\xi$ ({\em spectral parameter\/}), 
which appears in the equation for the spectral curve
\cite{Krichever,DHoker_Phong}. 
(iv) Independent conserved quantities \(Tr(L^k)\) to be obtained from a
Lax equation \(\dot{L}=[L,M]\) occur at such \(k= 1+  exponent\) of the
corresponding crystallographic root systems.
For the non-crystallographic root systems, they arise at  \(k=2,m\)
for the dihedral group \(I_2(m)\),  \(k=2,6,10\) for \(H_3\) and  
\(k=2,12, 20, 30\) for \(H_4\). These are the degrees at which Coxeter
invariant polynomials exist \cite{Coxeter_groups}.

\section{Lax pair and functional equations}
\label{consistency}
\setcounter{equation}{0}

Here we construct a Lax pair for the generalised Calogero-Moser models
in an operator form acting on an as yet unspecified vector space and
derive the necessary and sufficient conditions for the consistency 
of the Lax equations.
The Lax pair (\ref{LaxOpDef}) contains operators as well as functions
\(x_{|\rho|}(u,w)\), \(y_{|\rho|}(u,w)\), which are related to the
chosen potential (\ref{simppot})--(\ref{ratpot}).
The consistency of the Lax pair requires that \(x_{|\rho|}(u,w)\)
satisfy certain functional equations
(\ref{SimpleSumRule}), (\ref{fG_vanishes}), which are 
closely related with those required for the commutativity of
Dunkl operators \cite{Dunk,BFV}. Verification that the solutions
(\ref{Genl_A2_Soln}), (\ref{UntwistedB2Solns}), (\ref{TwistedB2Solns}),
 (\ref{TwistedG2Solns}),
(\ref{OddDihedralSoln}), (\ref{EvenDihedralSoln}) satisfy the functional
equations will be presented in the Appendix.

The operators appearing in the Lax pair for a generalised Calogero-Moser
model associated with a root system \(\Delta\) are naturally the 
reflection operators $\{\hat{s}_{\alpha},\,\alpha\in\Delta\}$ of the root
system.
They act on a set of $\mathbf{R}^{r}$ vectors 
\(\Gamma=\{\mu^{(k)}\in\mathbf{R}^{r},\ k=1,\ldots\}\), which is closed under
the action of the reflection group.
The totality of the vectors in \(\Gamma\) forms the representation
space \(\mathbf{V}\). A general construction of the representation
space, and
some explicit cases will be presented in the subsequent two sections.
Another set of operators $\{\hat{H}_{j},\,j=1,\ldots, r\}$ is necessary.
If  $\hat{H}_{j}$ acts on a vector $\mu^{(k)}\in\Gamma$, the $j$-th
component is returned:
\[
\hat{H}_{j}\mu^{(k)}=\mu^{(k)}_j\mu^{(k)}.
\]
These form the following operator algebra:
\begin{eqnarray}
   \label{OpAlgebra1}
   [\hat{H}_{j},\hat{H}_{k}]=0, \\ 
   \label{OpAlgebra2}
   [\hat{H}_{j},\hat{s}_{\alpha}] = \alpha_{j}
   (\alpha^{\vee}\!\!\cdot\hat{H})\hat{s}_{\alpha}, 
   \\ 
   \label{OpAlgebra3}
   \hat{s}_{\alpha}\hat{s}_{\beta}\hat{s}_{\alpha}
   =\hat{s}_{s_{\alpha}(\beta)}, 
   \\ 
   \label{OpAlgebra4}
   (\hat{s}_{\alpha}\hat{s}_{\beta})^{m(\alpha,\beta)}=1.
\end{eqnarray}
The first relation (\ref{OpAlgebra1}) implies that the 
operators $\{\hat{H}_{j},\,j=1,\ldots,
r\}$ form an abelian subalgebra and relations (\ref{OpAlgebra3}) and
(\ref{OpAlgebra4}) are just
those for the finite reflection group associated with the root system
$\Delta$.  The set of integers $m(\alpha,\beta)$ are those appearing
in the Coxeter relations 
(\ref{Coxeter_relations}) which characterise the reflection group.  

Next we describe the Lax pair and the corresponding Hamiltonian for
the generalised Calogero-Moser model for the root system $\Delta$.   The
Lax operators are  
\begin{eqnarray}
   \label{LaxOpDef}
   L &=& p\cdot\hat{H}+X,\qquad X=i\sum_{\rho\in\Delta_{+}}g_{|\rho|}
   \,\,(\rho^{\vee}\!\!\cdot\hat{H})\,x_{|\rho|}(\rho\cdot 
   q,(\rho^{\vee}\!\!\cdot\hat{H})\xi)\,\hat{s}_{\rho},
   \\ \nonumber
   M &=&
   i\sum_{\rho\in\Delta_{+}}g_{|\rho|}\,y_{|\rho|}
   (\rho\cdot q,(\rho^{\vee}\!\!\cdot\hat{H})\xi)\,\hat{s}_{\rho}.
\end{eqnarray}  
The function $y_{|\rho|}$ is defined by 
\begin{equation}
   y_{|\rho|}(u,w)\equiv {\partial\over {\partial u}} x_{|\rho|}(u,w).
\end{equation}
The variable $u$ takes care of the dynamical variable dependence and
$w$ is for the spectral parameter dependence. Furthermore, it is
required that
$x_{|\rho|}(u,w)$ is odd:
\begin{equation}
   x_{|\rho|}(-u,-w)=-x_{|\rho|}(u,w)
   \label{xodd}
\end{equation}
so that $L$ and 
$M$ are independent of the choice of positive roots $\Delta_{+}$.
This also implies that the sums in (\ref{LaxOpDef}) may be extended
to a sum over all roots if an additional factor of $1/2$ is included 
in front of
the sums since the summands are
even under $\rho\rightarrow -\rho$.  
The function $x_{|\rho|}(u,w)$ is assumed to have a simple pole at
$u=0$ with unit residue 
\begin{equation}
   \lim_{u\rightarrow 0}u\,x_{|\rho|}(u,w)=1.
   \label{unires}
\end{equation}
This condition 
is related with the unit strength of the repulsive potential near
the reflection hyperplane mentioned earlier.

Before proving the consistency of this Lax pair we calculate the
Hamiltonian and find the corresponding equations of motion.  The
Hamiltonian for the theory is defined in terms of a representation of
the 
operator $L$ of (\ref{LaxOpDef}) by
\begin{equation}
   \mathcal{H} = {1\over {2 C_{R}}}Tr(L^{2})
\end{equation}
where the constant $C_{R}$, which depends on the representation, is
defined by 
\begin{equation}
   Tr(\hat{H}_{j}\hat{H}_{k})=C_{R}\,\delta_{jk}.
\end{equation} 
The resulting Hamiltonian is then
\begin{equation}
   \mathcal{H} = {1\over 2}p^{2}+\sum_{\rho\in\Delta}{g_{|\rho|}^{2}\over
   |\rho|^{2}}\,V_{|\rho|}(\rho\cdot q)+ C,
\end{equation}
in which $C$ is independent of the dynamical variables $q$ and $p$ and
therefore unimportant for the classical theory. The potential functions
are given in  (\ref{simppot})--(\ref{ratpot}). The function  $x_{|\rho|}$
and the potential function
$V_{|\rho|}$ are related (except for the confining harmonic
potential (\ref{harmpot}), which has to be added separately) simply as
\begin{equation}
   \label{FactorIdentity}
   x_{|\rho|}(u,w)x_{|\rho|}(-u,w)=-V_{|\rho|}(u)+C_{|\rho|}(w),
\end{equation} 
{\it i.e.} the product gives a sum of a function of only $u$ and a
function of only $w$.  It is easy to show that all of the
functions $x_{|\rho|}(u,w)$ which lead to a consistent Lax equation,
(\ref{Genl_A2_Soln}), (\ref{UntwistedB2Solns}), (\ref{TwistedB2Solns}),
 (\ref{TwistedG2Solns}),
(\ref{OddDihedralSoln}), (\ref{EvenDihedralSoln}),  
satisfy this property.  

The equations of motion following from this Hamiltonian are 
\begin{eqnarray}
   \label{EqnsOfMotion}
   \dot{q}_{j} &=& {\partial\mathcal{H}\over{\partial p_{j}}} =  p_{j},\\
   \nonumber
   \dot{p}_{j} &=& -{\partial\mathcal{H}\over{\partial q_{j}}} 
   \\ \nonumber
   &=&-{\partial\over \partial q_{j}}\left[\sum_{\rho\in\Delta}
   {g_{|\rho|}^{2}\over |\rho|^{2}}V_{|\rho|}(\rho\cdot q)\right] \\ 
   &=&\sum_{\rho\in\Delta}{g_{|\rho|}^{2}\over |\rho|^{2}}\rho_{j}
   \left[y_{|\rho|}(\rho\cdot q,w)x_{|\rho|}(-\rho\cdot q,w)-x_{|\rho|}
   (\rho\cdot q,w)y_{|\rho|}(-\rho\cdot q,w)\right].
   \label{eqMot2}
\end{eqnarray}
Because of (\ref{FactorIdentity}) the last expression in 
(\ref{eqMot2}) is 
independent of $w$.

The Lax equation
\begin{equation}
   \label{LaxEquation}
   \dot{L}=[L,M]
\end{equation}
 may be divided into three parts as
\begin{eqnarray}
   \label{LaxEqn1}{d\over dt}X &=& [p\cdot\hat{H},M], \\
   \label{LaxEqn2}{d\over dt}(p\cdot\hat{H}) &=&
   [X,M]_{\mbox{\scriptsize\it diagonal}},\\
   \label{LaxEqn3}0 
   &=& [X,M]_{\mbox{\scriptsize\it off-diagonal}}.
\end{eqnarray}
We next prove, in turn, that each of these equations is consistent
with the equations of motion (\ref{EqnsOfMotion}).

The left-hand side of (\ref{LaxEqn1}) is
\begin{equation}
   {d\over dt}X = i\sum_{\rho\in\Delta_{+}}g_{|\rho|}
   \,(\rho^{\vee}\!\!\cdot\hat{H})\,y_{|\rho|}(\rho\cdot  
   q,(\rho^{\vee}\!\!\cdot\hat{H})\xi)\,
    (\rho\cdot\dot{q})\,
   \hat{s}_{\rho}
\end{equation}
and the right-hand side is
\begin{eqnarray}
   [p\cdot\hat{H},M] &=& \Bigl[p\cdot\hat{H},i\sum_{\rho\in\Delta_{+}}
   g_{|\rho|}\,y_{|\rho|}(\rho\cdot
   q,(\rho^{\vee}\!\!\cdot\hat{H})\xi)\,\hat{s}_{\rho}\Bigr], 
   \nonumber\\ \nonumber
   &=&i\sum_{\rho\in\Delta_{+}}g_{|\rho|}\,y_{|\rho|}(\rho\cdot
   q,(\rho^{\vee}\!\!\cdot\hat{H})\xi)\,[p\cdot\hat{H},\hat{s}_{\rho}], 
   \\ \nonumber
   &=& i\sum_{\rho\in\Delta_{+}}g_{|\rho|}\,y_{|\rho|}(\rho\cdot
   q,(\rho^{\vee}\!\!\cdot\hat{H})\xi)\,(\rho^{\vee}\!\!\cdot\hat{H})
   \,(\rho\cdot
   p)\,\hat{s}_{\rho}, \\ 
   &=&{d\over dt}X. 
\end{eqnarray}
The third line follows from the commutation relation (\ref{OpAlgebra2})
and the
last line follows from the equation of motion $\dot{q}=p$.

The left-hand side of (\ref{LaxEqn2}), after using the equations of motion
(\ref{EqnsOfMotion}) is
\begin{equation}
   {d\over dt}(p\cdot\hat{H})= \sum_{\rho\in\Delta}{g_{|\rho|}^{2}\over
   |\rho|^{2}}(\rho\cdot\hat{H})\left[y_{|\rho|}(\rho\cdot q,w)
   x_{|\rho|}(-\rho\cdot q,w)-x_{|\rho|}(\rho\cdot q,w)y_{|\rho|}
   (-\rho\cdot q,w)\right].
\end{equation}
The summand
$x_{|\rho|}(u,w)y_{|\rho|}(-u,w)-x_{|\rho|}(-u,w)y_{|\rho|}(u,w)$ is
independent of the parameter $w$. The commutator \([L,M]\) reads
\begin{eqnarray}
   [X,M]   \nonumber
   &=&-\Bigl[\sum_{\rho\in\Delta_{+}}
   g_{|\rho|}\,\,(\rho^{\vee}\!\!\cdot\hat{H})\,x_{|\rho|}(\rho\cdot
q,(\rho^{\vee}\!\!\cdot\hat{H})\xi)\, 
  \hat{s}_{\rho}, 
   \sum_{\sigma\in\Delta_{+}}g_{|\sigma|}\,y_{|\sigma|}(\sigma\cdot 
   q,(\sigma^{\vee}\!\!\cdot\hat{H})\xi)\hat{s}_{\sigma}\Bigr],\\
   \nonumber
   &=&-\sum_{\rho,\sigma\in\Delta_{+}}g_{|\rho|}g_{|\sigma|}
   \Bigl[(\rho^{\vee}\!\!\cdot\hat{H})\,x_{|\rho|}(\rho\cdot
   q,(\rho^{\vee}\!\!\cdot\hat{H})\xi)\,y_{|\sigma|}(\sigma\cdot
   q,(s_{\rho}(\sigma)^{\vee}\!\!\cdot\hat{H})\xi)
   \hat{s}_{\rho}\hat{s}_{\sigma} 
   \\ 
   &&\qquad\qquad -y_{|\sigma|}(\sigma\cdot
   q,(\sigma^{\vee}\!\!\cdot\hat{H})\xi)\,(s_{\sigma}
   (\rho)^{\vee}\!\!\cdot\hat{H})
   \,x_{|\rho|}(\rho\cdot
   q,(s_{\sigma}(\rho)^{\vee}\!\!\cdot\hat{H})\xi)\,\hat{s}_{\sigma}
   \hat{s}_{\rho}\Bigr].
  \label{xmcomm}
\end{eqnarray}
Since \(\hat{s}_{\rho}\hat{s}_{\sigma}\) and 
\(\hat{s}_{\sigma}\hat{s}_{\rho}\) are 
rotations (except for \(\rho=\sigma\), \(\hat{s}_{\rho}^2=1\)) 
they do not leave any
real vectors in the rotation plane invariant. Thus \([X,M]\) is decomposed
into the diagonal (\(\rho=\sigma\)) and the off-diagonal 
(\(\rho\neq\sigma\)) parts. The diagonal part gives the equation of motion
\begin{eqnarray}
   [X,M]_{\mbox{\scriptsize\it diag.}}  
   &=& \sum_{\rho\in\Delta_{+}}{g_{|\rho|}^{2}
    }(\rho^{\vee}\!\!\cdot \hat{H})\Bigl[y_{|\rho|}(\rho\cdot
   q,(\rho^{\vee}\!\!\cdot\hat{H})\xi)\,x_{|\rho|}(-\rho\cdot
   q,(\rho^{\vee}\!\!\cdot\hat{H})\xi) \nonumber\\ 
   &&\qquad\qquad\qquad\ \ -x_{|\rho|}(\rho\cdot
   q, (\rho^{\vee}\!\!\cdot\hat{H})\xi)\,y_{|\rho|}(-\rho\cdot
   q,(\rho^{\vee}\!\!\cdot\hat{H})\xi)\Bigr], \\ 
   &=& {d\over dt}(p\cdot\hat{H}).
   \label{diageqmot}
\end{eqnarray}

Finally, (\ref{LaxEqn3}) will lead to the functional equations which
must be
satisfied by the functions $x_{|\rho|}(u,w)$ in the Lax pair. 
Relabeling the roots in the second summation of (\ref{xmcomm}) 
gives the general
operator form for the functional equation  
\begin{eqnarray}
   0&=&[X,M]_{\mbox{\scriptsize\it off-diag.}}\nonumber\\ \nonumber
   &=&\sum_{\rho\neq\sigma\in\Delta_{+}}g_{|\rho|}g_{|\sigma|}
   \Bigl[(\rho^{\vee}\!\!\cdot\hat{H})x_{|\rho|}(\rho\cdot
   q,(\rho^{\vee}\!\!\cdot\hat{H})\xi)y_{|\sigma|}(\sigma\cdot
   q,(s_{\rho}(\sigma)^{\vee}\!\!\cdot\hat{H})\xi)\nonumber\\ 
   &&\qquad\qquad\quad -(s_{\rho}(\sigma)^{\vee}\!\!\cdot\hat{H})
   y_{|\rho|}(\rho\cdot q,(\rho^{\vee}\!\!\cdot\hat{H})\xi)
   x_{|\sigma|}(\sigma\cdot
   q,(s_{\rho}(\sigma)^{\vee}\!\!\cdot\hat{H})\xi)\Bigr]
   \hat{s}_{\rho}\hat{s}_{\sigma}.
   \label{GenlSumRule}
\end{eqnarray}

This shows that the consistency condition involving all of the roots
is decomposed into a sum of two-dimensional ones corresponding
to a fixed rotation $\hat{R}_{\psi}\equiv\hat{s}_{\rho}\hat{s}_{\sigma}$
 in
each two-dimensional plane.
Since the coefficient of
$\hat{R}_{\psi}\equiv\hat{s}_{\rho}\hat{s}_{\sigma}$ 
in this equation must separately 
vanish, this may be
decomposed into the functional equations 
\begin{eqnarray}
   \label{TwoDSumRules}
   0&=&\sum_{\rho\neq\sigma\in\Phi_{+},\,R_{\psi}=s_{\rho}s_{\sigma}}
   \!\!g_{|\rho|}g_{|\sigma|}
   \Bigl[(\rho^{\vee}\!\!\cdot\mu)x_{|\rho|}(\rho\cdot
   q,(\rho^{\vee}\!\!\cdot\mu)\xi)  
   y_{|\sigma|}(\sigma\cdot
   q,(s_{\rho}(\sigma)^{\vee}\!\!\cdot\mu)\xi)\nonumber\\ 
   &&\hspace{3cm}-(s_{\rho}(\sigma)^{\vee}\!\!\cdot\mu)
   y_{|\rho|}(\rho\cdot q,(\rho^{\vee}\!\!\cdot\mu)\xi) 
   \, x_{|\sigma|}(\sigma\cdot
   q,(s_{\rho}(\sigma)^{\vee}\!\!\cdot\mu)\xi)\Bigr],
\end{eqnarray}
in which \(\mu\) is a generic vector in \(\Gamma\).
This equation must be satisfied for a fixed rotation 
$R_{\psi}=s_{\rho}s_{\sigma}$ and all roots appearing in it are in the
two-dimensional sub-root system $\Phi=\{\kappa,\,\kappa\in(\Delta\cap
span(\rho,\sigma)\}$ with positive roots 
$\Phi_{+}\equiv \Phi\cap \Delta_{+}$.
The only possible two-dimensional root systems $\Phi$ are 
$A_{1}\times A_{1}$,
$A_{2}$, $B_{2}$, $G_{2}$, and $I_{2}(m)$.  
 Table 1 shows the
two-dimensional sub-root systems appearing in the root systems
of finite reflection groups.
The $A_{1}\times A_{1}$ root system has been omitted since its 
corresponding functional equation 
is trivially satisfied for any function 
and therefore does not give any constraint on the
functions $x_{|\rho|}(u,w)$. It should be stressed that the
functional equations are determined by the two-dimensional sub-root
systems only and not by where they are embedded in the entire root system.
Thus, as seen from Table 1, each function \(x_{|\rho|}\) must satisfy at most
two functional equations, except for the models based on the dihedral groups.
\begin{table}
\centering
   \begin{tabular}{|c|c|} \hline \label{SubrootTbl}
   Root System & Sub-root Systems \\ \hline
   $A_{r},\,r>1$ & $A_{2}$ \\ \hline
   $B_{r},\,r\ge2$ & $A_{2}$,\,$B_{2}$ \\ \hline
   $C_{r},\,r\ge2$ & $A_{2}$,\,$B_{2}$ \\ \hline
   $D_{r},\,r>3$ & $A_{2}$ \\ \hline
   $BC_{r},\,r\ge2$ & $A_{2}$,\,$B_{2}$ \\ \hline
   $E_{6}$,$E_{7}$,\,$E_{8}$ & $A_{2}$ \\ \hline
   $F_{4}$ & $A_{2}$,\,$B_{2}$ \\ \hline
   $G_{2}$ & $A_{2}$,\,$G_{2}$ \\ \hline
   $I_{2}(m)$ & $I_{2}(k)^\dagger$\\ \hline
   $H_{3}$ & $A_{2}$,\,$I_{2}(5)$ \\ \hline
   $H_{4}$ & $A_{2}$,\,$I_{2}(5)$ \\ \hline
   \end{tabular}
\caption{Two-dimensional sub-root systems.  
$A_{1}\times A_{1}$ is not included.
$\dagger$: $k$ divides  $m$.}
\end{table}

The functional equation in (\ref{TwoDSumRules}) 
may be further simplified to 
\begin{equation}
   \label{SimpleSumRule}
   0 = \mu\cdot R_{\psi/2}\ {\partial\over {\partial q}}f(q,\mu),
\end{equation}
in which 
\begin{equation}
   \label{GenlfDef}
   f(q,\mu) =
   \sum_{\rho\neq\sigma\in\Phi_{+},\,R_{\psi}=
   s_{\rho}s_{\sigma}}g_{|\rho|}  
   \,g_{|\sigma|}\,{(\sigma^{\vee}\!\!\cdot\rho)\over
   {|\sigma^{\vee}\!\!\cdot\rho|}}\,
   x_{|\rho|}(\rho\cdot 
   q,(\rho^{\vee}\!\!\cdot\mu)\xi)\,x_{|\sigma|}(\sigma\cdot
   q,(s_{\rho}(\sigma)^{\vee}\!\!\cdot\mu)\xi)
\end{equation}
and $R_{\psi/2}$ is a  rotation by an angle
$\psi/2$ with $R_{\psi}=s_{\rho}s_{\sigma}$.
It  has the following action on any pair of 
roots $\rho, \sigma\in\Phi_{+}$ which satisfy 
$R_{\psi}=s_{\rho}s_{\sigma}$
\begin{eqnarray}
   \label{RthetaAction}
   R_{\psi/2}\ \rho&=&-{|\rho|\over|\sigma|}\,
   {(\sigma^{\vee}\!\!\cdot\rho)\over
   {|\sigma^{\vee}\!\!\cdot\rho|}}\,s_{\rho}(\sigma), \\
   \nonumber
   R_{\psi/2}\ \sigma&=&{|\sigma|\over|\rho|}\,
   {(\sigma^{\vee}\!\!\cdot\rho)\over
   {|\sigma^{\vee}\!\!\cdot\rho|}}\,\rho.
\end{eqnarray}
Using these relations, the simplified form of the functional equation
(\ref{SimpleSumRule}) may be proven.  First substitute the definition
of $f(q,\mu)$ in (\ref{GenlfDef}) into (\ref{SimpleSumRule}) to obtain
\begin{eqnarray}
   0&=&\sum_{\rho\neq\sigma\in\Phi_{+},\,r=s_{\rho}s_{\sigma}}
   \!\!\!g_{|\rho|}g_{|\sigma|}{(\sigma^{\vee}\!\!\cdot\rho)\over
   {|\sigma^{\vee}\!\!\cdot\rho|}}\,\Biggl[(\mu\cdot 
   R_{\psi/2}\ \rho)y_{|\rho|}(\rho\cdot q,(\rho^{\vee}\!\!\cdot
   \mu)\xi)\,x_{|\sigma|}(\sigma\cdot
   q,(s_{\rho}(\sigma)^{\vee}\!\!\cdot\mu)\xi) \nonumber\\ 
   &&\hspace{3.5cm} +\ (\mu\cdot R_{\psi/2}\
   \sigma)x_{|\rho|}(\rho\cdot
   q,(\rho^{\vee}\!\!\cdot\mu)\xi)\,y_{|\sigma|}(\sigma\cdot
   q,(s_{\rho}(\sigma)^{\vee}\!\!\cdot\mu)\xi)\Biggr].
\end{eqnarray}
Since the coupling constants $g_{|\alpha|}$ are arbitrary and 
are constants on the 
orbits they may be rescaled as $g_{|\alpha|}\rightarrow
2g_{|\alpha|}/|\alpha|$ for all $\alpha\in\Phi$.  Performing this
rescaling of the coupling constants and using 
the action of $R_{\psi/2}$ in
(\ref{RthetaAction}), the previous equation becomes
\begin{eqnarray}
   \nonumber
   0&=&\hspace{-0.5cm}\sum_{\rho\neq\sigma\in\Phi_{+},
   \,r=s_{\rho}s_{\sigma}}\!\!\!g_{|\rho|}
   g_{|\sigma|}{2\over{|\rho||\sigma|}}\Biggl[{|\rho|\over|\sigma|}
   (-s_{\rho}(\sigma)\cdot\mu)y_{|\rho|}(\rho\cdot 
   q,(\rho^{\vee}\!\!\cdot\mu)\xi)\,x_{|\sigma|}(\sigma\cdot
   q,(s_{\rho}(\sigma)^{\vee}\cdot\mu)\xi) \\ 
   &&\hspace{3.6cm} +\
{|\sigma|\over|\rho|}(\rho\cdot\mu)x_{|\rho|}(\rho\cdot
   q,(\rho^{\vee}\cdot\mu)\xi)\,y_{|\sigma|}(\sigma\cdot
   q,(s_{\rho}(\sigma)^{\vee}\cdot\mu)\xi)\Biggr].
\end{eqnarray}
This is the same as the earlier form of the functional equation in
(\ref{TwoDSumRules}), after using the definition
$\alpha^{\vee}=2\alpha/|\alpha|^{2}$.

We next present the functional equations for the two-dimensional
root systems, $A_{2}$, $B_{2}$, $G_{2}$, and $I_{2}(m)$, and state
their solutions.  The proofs that the solutions satisfy the relevant
functional equations are contained in the Appendix.  

We first consider the functional equation  for the $A_{2}$ root system 
with simple
roots $\alpha$ and $\beta$. We choose   
$R_{2\pi/3}=s_{\alpha}s_{\beta}$ and the functional equation 
(\ref{GenlSumRule}) becomes
\begin{equation}
   \label{A2SumRule}
   0 = \mu\cdot R_{\pi/3}\ {\partial\over{\partial
   q}}f_{A_{2}}(q,\mu)
\end{equation}
in which $f_{A_{2}}(q,\mu)$ is defined by
\begin{eqnarray}
   f_{A_{2}}(q,\mu)=&&x((\alpha+\beta)\cdot q,
   (\alpha+\beta)^{\vee}\!\!\cdot
   \mu)\,x(\alpha\cdot q,-\beta^{\vee}\!\!\cdot\mu) \nonumber\\ 
   &+&x(\beta\cdot q,\beta^{\vee}\!\!\cdot\mu)\,x((\alpha+\beta)\cdot
   q,\alpha^{\vee}\!\!\cdot\mu) \nonumber\\ 
   &-&x(\alpha\cdot q,\alpha^{\vee}\!\!\cdot\mu)\,x(\beta\cdot
   q,(\alpha+\beta)^{\vee}\!\!\cdot\mu).
\end{eqnarray}
The subscripts on the function $x$ is omitted since all
roots are in the same orbit and hence only one function appears.  Also 
the spectral parameter $\xi$ is  absorbed into $\mu$ by redefinition:
$\mu\rightarrow\mu/\xi$.   We look for solutions to this functional 
equation, as well as the ones for other 
root systems, which are valid for arbitrary  vectors
$q$ and $\mu$. Therefore these solutions are valid in any representations.
In certain representations, such as the minimal and the root type
representations discussed in section \ref{minroot}, 
there is a larger class of solutions.
The functional equation arising from the rotation 
$R_{-2\pi/3}$ is equivalent to that given above.
A simple solution to (\ref{A2SumRule})  satisfying the residue 
condition (\ref{unires}) is
\begin{equation}
   \label{Genl_A2_Soln}
   x(u,w) = {\sigma(w-u)
   \over{\sigma(w)
   \sigma(u)}}
   \exp[b\,w\,u],
\end{equation}
in which  $b$ is an arbitrary complex parameter.  
The potential function in the resulting 
Hamiltonian (\ref{CMHamiltonian}), 
$V(u)=\wp(u)$, 
is the Weierstrass
elliptic function with the set of primitive periods 
$2\omega_{1}$ and $2\omega_{3}$.  
The limits as one or both periods diverge yield the 
potential functions $a^2/\sin^{2}(au)$ and $a^2/\sinh^{2}(au)$ or 
$1/u^{2}$, respectively \cite{Our_CM_Papers}.

Functional equation (\ref{TwoDSumRules}) for the 
$B_{2}$ root system may be
written in terms of the short and long simple roots, $\alpha$ and
$\beta$, respectively.  We choose   $R_{\pi/2}=s_{\alpha}s_{\beta}$
in (\ref{GenlSumRule})  and the resulting functional equation is 
\begin{equation}
   \label{B2SumRule}
   0 = \mu\cdot R_{\pi/4}\ {\partial\over{\partial
   q}}f_{B_{2}}(q,\mu)
\end{equation}
in which
\begin{eqnarray}
   f_{B_{2}}(q,\mu) =&-& x_{S}(\alpha\cdot
   q,\alpha^{\vee}\!\!\cdot\mu)\,x_{L}
   (\beta\cdot q,(2\alpha+\beta)^{\vee}\!\!\cdot\mu)
   \nonumber\\ \nonumber
   &+&x_{S}((\alpha+\beta)\cdot
   q,(\alpha+\beta)^{\vee}\!\!\cdot\mu)\,x_{L}((2\alpha+\beta)\cdot
   q,-\beta^{\vee}\!\!\cdot\mu) \\ \nonumber
   &+&x_{S}(\alpha\cdot
   q,-(\alpha+\beta)^{\vee}\!\!\cdot\mu)\,x_{L}((2\alpha+\beta)\cdot
   q,(2\alpha+\beta)^{\vee}\!\!\cdot\mu) \\ 
   &+&x_{S}((\alpha+\beta)\cdot
   q,\alpha^{\vee}\!\!\cdot\mu)\,
    x_{L}(\beta\cdot q,\beta^{\vee}\!\!\cdot\mu).
   \label{fBfunc}
\end{eqnarray}
In this case there are two sets of elliptic function solutions for the
long and short root functions, $x_{L}$ and $x_{S}$, respectively.  The
solutions along with the corresponding potential functions are
\begin{eqnarray}
   \label{UntwistedB2Solns}
   x_{L}(u,w) =x_{S}(u,w) &=& {\sigma(w-u)
   \over{\sigma(w)
   \sigma(u)}}
   \exp[b\,w\,u], \\ \nonumber
   V_{L}(u)=V_{S}(u)&=&\wp(u),
\end{eqnarray}
for the untwisted solution
and 
\begin{eqnarray}
   x_{L}(u,w)&=& {\sigma(w-u)
   \over{\sigma(w)
   \sigma(u)}}
   \exp[b\,w\,u], \nonumber\\ 
   x_{S}(u,w)&=& {\sigma(w/2-u|\{\omega_{1},2\omega_{3}\})
   \over{\sigma(w/2|\{\omega_{1},2\omega_{3}\})
   \sigma(u|\{\omega_{1},2\omega_{3}\})}}
   \exp[(b+{e_{1}\over 2})\,w\,u], \nonumber\\ 
   &=& {{x_{L}(u,w/2)x_{L}(u+\omega_{1},w/2)}\over
   x_{L}(\omega_{1},w/2)},\label{TwistedB2Solns}\\ \nonumber
   V_{L}(u) &=& \wp(u), \qquad\qquad
   V_{S}(u) = \wp(u|\{\omega_{1},2\omega_{3}\}),
\end{eqnarray}
for the twisted solution.
Here $e_{1}\equiv \wp(\omega_{1})$ and $b$ is 
an arbitrary complex constant.

Next we consider the functional equation 
(\ref{TwoDSumRules}) for the $G_{2}$
root system with short and long simple roots $\alpha$ and $\beta$,
respectively. We choose  $R_{\pi/3}=s_{\alpha}s_{\beta}$ in
(\ref{TwoDSumRules})  and
the functional equation is
\begin{equation}
   \label{G2SumRule}
   0 = \mu\cdot R_{\pi/6}\ {\partial\over{\partial
   q}}f_{G_{2}}(q,\mu),
\end{equation}
in which
\begin{eqnarray}
   f_{G_{2}}(q,\mu)=&-&x_{S}(\alpha\cdot
   q,\alpha^{\vee}\!\!\cdot\mu)\,x_{L}
   (\beta\cdot q,(3\alpha+\beta)^{\vee}\!\!\cdot\mu)
   \nonumber\\ \nonumber
   &+&x_{S}(\alpha\cdot
   q,-(2\alpha+\beta)^{\vee}\!\!\cdot\mu)\,x_{L}((3\alpha+\beta)\cdot
   q,(3\alpha+\beta)^{\vee}\!\!\cdot\mu) \\ \nonumber
   &+&x_{S}((2\alpha+\beta)\cdot
   q,(2\alpha+\beta)^{\vee}\!\!\cdot\mu)\,x_{L}((3\alpha+\beta)\cdot
   q,-(3\alpha+2\beta)^{\vee}\!\!\cdot\mu) \\ \nonumber
   &+&x_{S}((2\alpha+\beta)\cdot
   q,-(\alpha+\beta)^{\vee}\!\!\cdot\mu)\,x_{L}((3\alpha+2\beta)\cdot
   q,(3\alpha+2\beta)^{\vee}\!\!\cdot\mu) \\ \nonumber
   &+&x_{S}((\alpha+\beta)\cdot
   q,(\alpha+\beta)^{\vee}\!\!\cdot\mu)\,x_{L}((3\alpha+2\beta)\cdot
   q,-\beta^{\vee}\!\!\cdot\mu)\\ 
   &+&x_{S}((\alpha+\beta)\cdot 
   q,\alpha^{\vee}\!\!\cdot\mu)\,
   x_{L}(\beta\cdot q,\beta^{\vee}\!\!\cdot\mu).
\end{eqnarray}
As before, $x_{S}$ and $x_{L}$ are the functions for short and long
roots, respectively.
The elliptic function solutions to (\ref{G2SumRule}) along with the 
corresponding potential functions $V_{S}(u)$ and $V_{L}(u)$ 
are the untwisted ones in 
(\ref{UntwistedB2Solns}) and the following twisted ones
\begin{eqnarray}
   x_{L}(u,w)&=& {\sigma(w-u)
   \over{\sigma(w)
   \sigma(u)}}
   \exp[b\,w\,u], \nonumber\\ 
   x_{S}(u,w)&=& {\sigma(w/3-u|\{2\omega_{1}/3,2\omega_{3}\})
   \over{\sigma(w/3|\{2\omega_{1}/3,2\omega_{3}\})
   \sigma(u|\{2\omega_{1}/3,2\omega_{3}\})}}
   \exp[(b+{2\over 3}\wp(2\omega_{1}/3))\,w\,u], 
   \nonumber\\ 
   &=& {{x_{L}(u,w/3)x_{L}(u+2\omega_{1}/3,w/3)x_{L}
   (u+4\omega_{1}/3,w/3)}\over
   {x_{L}(2\omega_{1}/3,w/3)x_{L}(4\omega_{1}/3,w/3)}}\exp[b\,w\,u],
   \label{TwistedG2Solns}\\
   V_{L}(u) &=& \wp(u), \qquad\qquad
   V_{S}(u) = \wp(u|\{2\omega_{1}/3,2\omega_{3}\}).
   \nonumber
\end{eqnarray}
Here again, $b$ is an arbitrary complex
parameter.  The short and long root functions also satisfy the
$A_{2}$ functional equation (\ref{A2SumRule}) separately, as expected.

Finally, we consider the functional equation (\ref{TwoDSumRules}) for the
$I_{2}(m)$ root system.  The $2m$ roots are labelled in order
of increasing angle by $\{\alpha_{1},\ldots,\alpha_{2m}\}$ with
$\Delta_{+}=\{\alpha_{j},j=1,\ldots,m\}$ and
$\alpha_{j+m}=-\alpha_{j}$ for $j=1,\ldots,m$.  For
example, choosing all roots to have the same length $|\alpha_{j}|=1$, 
a possible basis is given in (\ref{DihedralBasis}).  Unlike the
crystallographic root systems, there is, in general, more than one
functional equation. The functional equation
(\ref{TwoDSumRules})  corresponding to 
$R_{\psi}=s_{\rho}s_{\sigma}$ with $\psi=2\pi N/m,\,N=1,\ldots,[m/2]$ is
\begin{equation}
   \label{I2mSumRule}
   0 = \mu\cdot R_{{N\pi}/m}\ {\partial\over{\partial
   q}}f_{I_{2}(m)}^{N}(q,\mu)
\end{equation}
with
\begin{equation}
   \label{I2mfDef}
   f_{I_{2}(m)}^{N}(q,\mu)=\sum_{j=1}^{m}g_{|j|}\,g_{|j+N|}\,x_{|j+N|}
   (\alpha_{j+N}\cdot
   q,\alpha_{j+N}^{\vee}\!\!\cdot\mu)\,x_{|j|}(\alpha_{j}\cdot
   q,-\alpha_{j+2N}^{\vee}\!\!\cdot\mu).
\end{equation}
In contrast to the previous cases, the the coupling constants 
are included in the functional equation since they do not factor
out, in general. 
 Because of the many functional equations (\ref{I2mSumRule})
to be satisfied,  only
rational  solutions are allowed in this case. 
For odd $m$, $I_{2}(m)$ roots are all in a single orbit and
only one coupling constant and function are possible. 
The solution to (\ref{I2mSumRule}) in this case is
\begin{equation}
   \label{OddDihedralSoln}
   x(u,w) = \left({1\over u}-{1\over w}\right)\exp[b\,u\,w].
\end{equation}
However, $I_{2}(m)$ for even $m$  has two orbits: one the set
$\{\alpha_{j}\}$ with odd $j$  and the other with  even $j$.  
It is possible to have two independent coupling constants and 
functions in this case.  
The corresponding functions are denoted $x_{O}$ and $x_{E}$, 
respectively and the
solution of (\ref{I2mSumRule}) is 
\begin{eqnarray}
   \label{EvenDihedralSoln}
   x_{O}(u,w)&=&\left({1\over u}-{1\over w}\right)\exp[b\,u\,w], 
   \\ \nonumber
   x_{E}(u,w)&=&\left({1\over u}-{c\over w}\right)\exp[b\,u\,w],
\end{eqnarray}
in which $b$ and $c$ are arbitrary complex constants.

For the other non-crystallographic root systems $H_3$ and $H_4$, 
the two-dimensional sub-root systems are $A_2$ and $I_2(5)$.
Thus the solution (\ref{OddDihedralSoln}) satisfies all of the functional
equations for the consistency of the Lax pair.

At the end of this section, let us show the Lax pair operator
formulation for
the rational potential with a confining harmonic force.
This applies, as before, to all of the root systems including the
non-crystallographic ones.
This is a simple generalisation of the Lax pairs given in
\cite{OP1}, which were constructed for the vector representations
of the classical root systems, \(A_r\), \(B_r\), \(C_r\), \(D_r\), and
\(BC_r\).

Let us start from the Lax operator for the rational potential 
without spectral parameter:
\begin{eqnarray}
   \label{LaxOprat}
   L &=& p\cdot\hat{H}+i\sum_{\rho\in\Delta_{+}}g_{|\rho|}
   \,\,(\rho^{\vee}\!\!\cdot\hat{H})\,x(\rho\cdot 
   q)\,\hat{s}_{\rho},
   \\ \nonumber
   M &=&
   i\sum_{\rho\in\Delta_{+}}g_{|\rho|}\,y
   (\rho\cdot q)\,\hat{s}_{\rho},\nonumber\\
   & & \hspace{-0.5cm}x(\rho\cdot q)={1\over{\rho\cdot q}},\quad 
   y(\rho\cdot q)=-{1\over{(\rho\cdot q)^2}}=-x(\rho\cdot q)^2.
\end{eqnarray}
It corresponds to the following Hamiltonian and the equations of motion:
\begin{eqnarray}
   Tr(L^2)&\propto&{\cal H}_r={1\over2}p^2+\sum_{\alpha\in\Delta} 
      {g_{|\alpha|}^{2}\over |\alpha|^{2}}{1\over{(\alpha\cdot q)^2}},
   \label{1strat}\\
   \dot{L}=[L,M]&\Leftrightarrow&\dot{q}=p,\quad
   \dot{p}=2\sum_{\alpha\in\Delta} 
      {g_{|\alpha|}^{2}\over |\alpha|^{2}}
   {\alpha\over{(\alpha\cdot q)^3}}.
\end{eqnarray}
If a confining harmonic potential is added to the Hamiltonian
\begin{equation}
   {\cal H}_{\omega}={1\over2}p^2+{1\over2}\omega^2q^2+
   \sum_{\alpha\in\Delta} 
      {g_{|\alpha|}^{2}\over |\alpha|^{2}}{1\over{(\alpha\cdot q)^2}},
   \label{1stratharm}
\end{equation}
the equations of motion read
\begin{equation}
   \dot{q}=p,\quad \dot{p}=-\omega^2q+2\sum_{\alpha\in\Delta} 
      {g_{|\alpha|}^{2}\over |\alpha|^{2}}
  {\alpha\over{(\alpha\cdot q)^3}}.
\end{equation}
It is elementary to see that the above equations 
can be written in an operator form
\begin{equation}
   \dot{L}=[L,M]-\omega^2Q, \quad \dot{Q}=p\cdot\hat{H}=L-X,
\end{equation}
in which \(L\) and \(M\) are the same as before 
(\ref{LaxOprat}) and \(Q\) is
defined by
\begin{equation}
   Q=q\cdot\hat{H}.
\end{equation}
It is easy to verify that
\begin{equation}
   [Q,M]=-X.
\end{equation}
This property is valid only for the rational potential without spectral
parameter.

Next let us introduce a pair of non-hermitian operators \(L^{\pm}\) by
\begin{equation}
   L^{\pm}=L\pm i\omega Q.
\end{equation}
Their time evolution equations read
\begin{eqnarray}
   \dot{L}^{\pm}&=&\dot{L}\pm i\omega \dot{Q}\nonumber\\
   &=&[L,M]-\omega^2 Q\pm i\omega(L-X)\nonumber\\
   &=&[L^{\pm}\mp i\omega Q,M]-\omega^2 Q \pm i\omega(L-X)
   \nonumber\\
   &=&[L^{\pm},M]-\omega^2 Q \pm i\omega(L^{\pm}\mp i\omega Q)
   \nonumber\\
   &=&[L^{\pm},M]\pm i\omega L^{\pm}.
   \label{omegaLM}
\end{eqnarray}
If we define hermitian operators \({\cal L}_1\) and \({\cal L}_2\)
\begin{equation}
   {\cal L}_1=L^+L^-,\quad {\cal L}_2=L^-L^+,
   \end{equation}
   they satisfy Lax-like equations
   \begin{equation}
   \dot{{\cal L}}_j=[{\cal L}_j,M],\quad j=1,2.
\end{equation}
\noindent From these we can construct conserved quantities
\begin{equation}
   Tr({\cal L}_j^k),\quad j=1,2,\quad k=1,2,\ldots,
\end{equation}
as before. It is elementary to check that the first conserved quantities
give the Hamiltonian (\ref{1stratharm})
\begin{equation}
   Tr({\cal L}_j)\propto {\cal H}_\omega,\quad j=1,2.
\end{equation}
This then completes the derivation of the Lax pairs for all of the
generalised  Calogero-Moser
models.

\section{Representations of the Lax operators}
\label{reps}
\setcounter{equation}{0}

We next consider representations of the 
operator algebra (\ref{OpAlgebra1}-\ref{OpAlgebra4}) for operators 
$\{\hat{H}_{j},\,j=1,\ldots r\}$ and
$\{\hat{s}_{\alpha},\,\alpha\in\Delta\}$ that appear in the
Calogero-Moser 
Lax pair (\ref{LaxOpDef}).  In general, the representation of an 
algebra consists of a
vector space $\mathbf{V}$ and a mapping from elements of the algebra to
$GL(\mathbf{V})$, {\it e.g.} $\hat{s}_{\alpha}\rightarrow
\mathcal{R}(\hat{s}_{\alpha})$.  
When a basis is chosen for $\mathbf{V}$ then 
$\mathcal{R}(\hat{s}_{\alpha})$ becomes a
matrix in $GL(d)$ where $d$ is the dimension of $\mathbf{V}$.  
First define a monomial 
$P_{\Omega}(t)=\prod_{j=1}^{N} (v^{(j)}\cdot t)$ 
in an auxiliary vector variable $t\in\mathbf{R}^{r}$
associated with a set $\Omega$ of vectors 
$\{v^{(j)}\in\mathbf{R}^{r}, j=1,\ldots
N\}$.  
The basis vectors of $\mathbf{V}$, for a representation 
$\mathcal{R}_{\Omega}$ of the algebra 
(\ref{OpAlgebra1}-\ref{OpAlgebra4}), are  
monomials 
resulting from the orbit 
of $P_{\Omega}(t)$ under the reflection group, {\it i.e.}
$P_{\Omega}(s_{\alpha}(t))$ with $\alpha\in\Delta
$\footnote{More generally,
the vector space for the representation is $Sym(W^{*})$, the symmetric 
algebra on the $r$-dimensional dual vector space $W^{*}$.}.  
All of the monomials
in $\mathbf{V}$ therefore have the same degree $N$.
A similar
representation for only the reflection generators $\hat{s}_{\alpha}$
with 
$\Omega$ a sub-root system of $\Delta$ was introduced by MacDonald for
irreducible representations of the Weyl groups \cite{MacDonald,Hawkins}. 
The action of the operators on the basis vectors in $\mathbf{V}$, is 
\begin{eqnarray} 
   \label{Algebra_Def}
   \hat{s}_{\alpha}\,\prod_{j=1}^{N}(v^{(j)}\cdot t) &=& 
   \prod_{j=1}^{N}(s_{\alpha}(v^{(j)})\cdot t), \\ \nonumber
  \hat{H}_{l}\,\prod_{j=1}^{N}(v^{(j)}\cdot t) &=& 
   \left[\sum_{j=1}^{N}v^{(j)}_{l}\right]
   \,\prod_{j=1}^{N}(v^{(j)}\cdot t). \\ \nonumber
\end{eqnarray}
Note that the representation matrices 
$\{\mathcal{R}(\hat{H}_{l}), l=1,\ldots,r\}$ 
are all diagonal in this basis.  

All of the previous versions of Lax pairs for the Calogero-Moser
models, namely the minimal and root type Lax pairs,
result from (\ref{LaxOpDef}) in a representation 
${\mathcal{R}}_{\Omega}$ with $\Omega$ a
single vector, {\it i.e.} (\ref{Algebra_Def}) 
with $N=1$ \cite{Our_CM_Papers}.  
In the case of $N=1$, the
vectors in $\mathbf{V}$, upon which the representation matrices act, 
may be denoted in a simple manner as 
$|\mu\rangle$ with
\begin{equation}
   \label{SimpleRepNotation}
   |\mu\rangle\equiv (\mu\cdot t).
\end{equation}
For the minimal type Lax pair, the representation has  
$|\mu\rangle\in \mathbf{V}$ with $\mu$ being a weight 
in the same Weyl orbit as
the highest weight of a minimal representation of the corresponding Lie
algebra.  For the root type Lax pair, 
the vectors $\mu$ in $\mathbf{V}$ are labelled by  
all of the roots for the simply-laced root systems and either the
short or long roots for the non-simply laced root systems.  
All of the 
$\mu$ which label the basis vectors of 
$\mathbf{V}$ are then in a single Weyl
orbit, as required.  
It will be demonstrated in section \ref{minroot} that
the Lax pairs in these representations agree with those given
in \cite{Our_CM_Papers}.

There is a simple relation between the geometry of the root system and 
the dimension of the ${\mathcal{R}}_{\Omega}$ representations for
$\Omega$ a 
single vector.
The vectors $\mu$ labelling the $|\mu\rangle\in \mathbf{V}$ 
are generated by the orbit of a single vector $\mu_{0}$ 
under the action of the finite reflection group.
The vector $\mu_{0}$ may also be assumed, without loss of generality, 
to be in the principal Weyl chamber or its boundary.  The dimension of 
the representation depends only on which reflection hyperplanes, if
any, contain the vector $\mu_{0}$.  
Let $\Phi_{0}$ be the set of
indices $j$ such that 
$\Phi_{0}\equiv\{j,\,\mu_{0}\cdot\alpha_{j}=0,\,\alpha_{j}\in\Pi\}$.
For crystallographic root systems $\Phi_{0}$ are the Dynkin labels of
$\mu_{0}$ which are equal to zero.
Then $s_{\alpha_{j}}(\mu_{0})=\mu_{0}$
for $j\in\Phi_{0}$ or the isotropy group $I_{0}$ of $\mu_{0}$ is
generated by reflections about simple roots $\alpha_{j}\in\Pi$ for which
$j\in\Phi_{0}$.  Therefore  the number of
elements in the orbit of $\mu_{0}$ under the reflection group $W$, and 
hence the dimension of the representation, is $D=|W|/|I_{0}|$.
Since the isotropy group is the direct product of the reflection groups
corresponding to the connected parts of the Coxeter diagram for the
original root system, after deleting those vertices corresponding to
the indices $j$ {\it not} contained in $I_{0}$, 
$|I_{0}|$ may be computed as the product of the
orders of the reflection groups of the corresponding Coxeter
sub-diagrams. 

As a first example, we calculate the dimensions of the minimal and root 
type representations of the algebra in
(\ref{OpAlgebra1}-\ref{OpAlgebra4}) and hence the Lax pairs for
$A_{5}$.  There are three
inequivalent minimal representations with $\mu_{0}=\Lambda^{(1)}$,
$\Lambda^{(2)}$, or $\Lambda^{(3)}$.    For $\mu_{0}=\Lambda^{(1)}$, the
vector representation, the isotropy group $I_{0}=W_{A_{4}}$ so 
$D=|W_{A_{5}}|/|W_{A_{4}}|=6!/5!=6$.
For the antisymmetric product of two
vector representations, $\mu_{0}=\Lambda^{(2)}$,  $I_{0}=W_{A_{1}}\times
W_{A_{3}}$ so we obtain
$D=6!/(2! 4!) = 15$.  
The antisymmetric product of three
vector representations,  $\mu_{0}=\Lambda^{(3)}$, 
$I_{0}=W_{A_{2}}\times W_{A_{2}}$ so we obtain
$D=6!/(3! 3!) = 40$.  Finally, for the root representation, choose 
$\mu_{0}=\Lambda^{(1)}+\Lambda^{(5)}$, which is  the highest
root. Then $I_{0}=W_{A_{3}}$ and $D=|W_{A_{5}}|/|W_{A_{3}}|=6!/4!=30$.
As an aside, we note that the highest dimensional representation of
the Lax pair for $A_{5}$  has
$\mu_{0}=a_1\Lambda^{(1)}+a_2\Lambda^{(2)}+a_3\Lambda^{(3)}
+a_4\Lambda^{(4)}+a_5\Lambda^{(5)}$ with all non-vanishing coefficients $a_{j},
j=1,\ldots, 5$.
Since the dimension of the representation depends only 
on which  Dynkin labels
are non-zero, these coefficients \(a_j\) 
may be chosen to be $1$, without loss
of generality. Then the isotropy group is trivial and we obtain 
$D=6!$, which has not been previously described.  In fact, even among the
representations $R_{\Omega}$ with $N=1$, most of the matrices for the Lax
pairs had not been described before. 

As a second example, let us evaluate the dimensions of some 
lower-dimensional Lax pairs for the \(E_8\) model. Let us take 
$\mu_{0}=\Lambda^{(1)}$ and 
$\mu_{0}=\Lambda^{(7)}$ corresponding to the two end points of
the two longer forks of the \(E_8\) Dynkin diagram.
For $\mu_{0}=\Lambda^{(1)}$ the isotropy group $I_{0}=W_{E_{7}}$
and $D=|W_{E_{8}}|/|W_{E_{7}}|=240$. This gives the root type Lax pair.
For $\mu_{0}=\Lambda^{(7)}$ the isotropy group $I_{0}=W_{D_{7}}$
and $D=|W_{E_{8}}|/|W_{D_{7}}|=2160$. 
This gives the second smallest Lax pair
for the $E_8$ model and its weights are a part of the {\bf 3875}
representation.

For the non-crystallographic root systems, the root type Lax pairs with
dimensions $2m$ for $I_2(m)$, 30 for $H_3$ and 120 for $H_4$ give
the smallest dimensional Lax pair matrices.

\section{Minimal and root type Lax pairs}
\label{minroot}
\setcounter{equation}{0}

As examples of representations of the Lax operators, we consider those 
representations that yield the minimal and root type Lax pairs
previously reported in Ref. \cite{Our_CM_Papers}. 
The functional equations associated with these representations are also
derived by restricting those given in section \ref{consistency}.
These are the same as those given in our previous papers, which we denote
as  sum rules. In this section we consider only the
crystallographic root systems. 

Minimal type Lax pairs are associated with minimal representations of the
Lie algebras. 
All of the fundamental representations of $A_{r}$,  the spinor
representation of
$B_{r}$, the vector representation of $C_{r}$, and the vector, spinor
and anti-spinor representations of
$D_{r}$, the  {\bf 27} and 
\({\bf \overline{27}}\) of \(E_6\)
and the {\bf 56} of
\(E_{7}\) are the minimal representations. 
All weights in a minimal representation are in a single Weyl orbit.  
The vectors $|\mu\rangle\in \mathbf{V}$ of the representation are 
$\{|\mu\rangle,\ \mu\in\Sigma_{min}\}$, 
in which $\Sigma_{min}$ is the set of weights of the minimal
representation.       They are characterised by the condition
\begin{equation}
   \label{MinRepCondition}
   \rho^{\vee}\!\!\cdot\mu=\{0,\pm1\},\qquad
   \rho\in\Delta,\quad \mu\in\Sigma_{min}.
\end{equation}

On the other hand the  vectors $|\mu\rangle\in \mathbf{V}$ of the
root type Lax pairs are 
$\{|\alpha\rangle,\,\alpha\in\Delta\}$ for a simply-laced root
system 
$\Delta$. 
The set of basis vectors of the two possible root type
representations for a Lax pair associated with a non-simply laced root
system are 
$\{|\alpha\rangle,\,\alpha\in\Delta_{S}\}$, the set of short roots,
and 
$\{|\alpha\rangle,\,\alpha\in\Delta_{L}\}$, the set of long roots. 
Let us collectively denote by \(\Delta_R\) the set of basis vectors
of various root type representations:
\[
\Delta_R=\cases{\Delta, &\mbox{for simply laced models, \quad\ \ \ all
roots},\cr
\Delta_L, &\mbox{for non-simply laced models, long roots},\cr
\Delta_S, &\mbox{for non-simply laced models, short roots}.\cr}
\]
The roots are characterised by the condition
\begin{equation}
   \label{rootRepCondition}
   \rho^{\vee}\!\!\cdot\alpha=\{0,\,\pm1,\,\pm2\},\qquad
   \rho,\alpha\in\Delta,
\end{equation}
except for the \(G_2\) case in which \(\pm3\) are also possible.

The fact that the eigenvalues of the operator 
\(\rho^{\vee}\!\!\cdot\hat{H}\)
are restricted to these values (\ref{MinRepCondition}),
(\ref{rootRepCondition}) simplifies the representation of the Lax pair
operator (\ref{LaxOpDef}) drastically. Especially at the zero eigenvalue
of \(\rho^{\vee}\!\!\cdot\hat{H}\) the functions \(x_{|\rho|}\) and 
\(y_{|\rho|}\) take the following simple forms:
\begin{equation}
   \label{x_w_Limit}
   \lim_{w\rightarrow 0}w\,x_{|\rho|}(u,w\xi)=
   \cases{-{1/\xi},\quad \mbox{untwisted},\cr
       -{2/\xi},\quad \mbox{twisted short roots},\cr
       -{3/\xi},\quad \mbox{twisted short roots},\ G_2,\cr}
\end{equation}
\begin{equation}
   \label{y_w_Limit}
   \lim_{w\rightarrow 0}y_{|\rho|}(u,w)=-V_{|\rho|}(u)+D_{|\rho|},
\end{equation}
in which \(D_{|\rho|}\) are constants possibly depending on the
orbits, that is, \(D_{L}\) and \(D_{S}\) for the non-simply laced cases.

With the basis vectors of $\mathbf{V}$, upon which 
the representation matrices act, 
labelled by a single vector, as
in (\ref{SimpleRepNotation}), the matrix elements of the minimal and root
type Lax
pairs are
\begin{eqnarray}
   \label{minLaxMatrixElements}
   L|\nu\rangle  &=&
\sum_{\mu\in\Sigma_{min}}\mathcal{R}(L)_{\mu\nu}|\mu\rangle ,
\qquad 
   M|\nu\rangle = \sum_{\mu\in\Sigma_{min}}\mathcal{R}(M)_{\mu\nu}|
\mu \rangle,\qquad
   \mu, \nu\in\Sigma_{min},\\
L|\beta\rangle  &=&\ \sum_{\alpha\in\Delta_{R}}
\mathcal{R}(L)_{\alpha\beta} |\alpha \rangle, \qquad 
M|\beta\rangle   =\ \sum_{\alpha\in\Delta_{R}}
 \mathcal{R}(M)_{\alpha\beta}|\alpha\rangle ,\qquad\
   \alpha, \beta\in\Delta_R, 
 \label{rootLaxMatrixElements}
\end{eqnarray}
where $L$ and $M$ are the Lax operators defined in (\ref{LaxOpDef}).
Hereafter we adopt simplified notation \(L_{\mu\nu}\) 
(\(L_{\alpha\beta}\)) and
\(M_{\mu\nu}\) (\(M_{\alpha\beta}\)) for the matrix
elements \(\mathcal{R}(L)_{\mu\nu}\)  (\(\mathcal{R}(L)_{\alpha\beta}\)
) and
\(\mathcal{R}(M)_{\mu\nu}\) (\(\mathcal{R}(M)_{\alpha\beta}\) ). For an
arbitrary function $f(u)$ we have the following matrix element for the minimal
type
\begin{equation}
   f(\rho^{\vee}\!\!\!\cdot\hat{H})\hat{s}_{\rho}|\nu\rangle
=\sum_{\mu\in\Sigma_{min}}\biggl\{
   f(-1)\delta_{\mu-\nu,-\rho}+f(0)\delta_{\mu,\nu}
\delta_{\rho\cdot\mu,0}
   +f(1)\delta_{\mu-\nu,\rho}\biggr\}|\mu\rangle,
   \label{fvalmin}
\end{equation}
and for the root type (except for the \(G_2\) case)
\begin{eqnarray}
   f(\rho^{\vee}\!\!\cdot\hat{H})\hat{s}_{\rho}|\beta\rangle &=&
\sum_{\alpha\in\Delta_{R}}\biggl\{
   f(-2)\delta_{\alpha-\beta,-2\rho}+f(-1)
\delta_{\alpha-\beta,-\rho}
   +f(0)\delta_{\alpha,\beta}\delta_{\rho\cdot\alpha,0}
\nonumber\\
   &&\qquad\qquad\qquad + f(1)\delta_{\alpha-\beta,\rho}
   +f(2)\delta_{\alpha-\beta,2\rho}\biggr\}|\alpha\rangle.
\label{fvalroot}
\end{eqnarray}
The \(G_2\) case can be handled in a similar way.

By combining (\ref{x_w_Limit}), (\ref{y_w_Limit}) and (\ref{fvalmin}),
(\ref{fvalroot}) it is straightforward to derive the matrix
representations of the minimal and root type Lax pairs. For the minimal
Lax pair we obtain
\begin{eqnarray}
   L_{\mu\nu}&=&p\cdot\mu\,\delta_{\mu,\nu}+i
   \sum_{\rho\in\Delta}g_{|\rho|} x_{|\rho|}(\rho\cdot
   q,\xi)\delta_{\mu-\nu,\rho}+A_m\delta_{\mu,\nu}, 
   \label{lmin}\\
   M_{\mu\nu}&=&i\left[\sum_{\rho\in\Delta,
\,\rho^{\vee}\!\!\cdot\mu=1}
   g_{|\rho|}V_{|\rho|}(\rho\cdot q)\right]\delta_{\mu,\nu}
     + i \sum_{\rho\in\Delta}g_{|\rho|}y_{|\rho|}(\rho\cdot
   q,\xi)\delta_{\mu-\nu,\rho} +B_m\delta_{\mu,\nu},
   \label{mmin}
\end{eqnarray}
in which \(A_m\) is a constant and \(B_m\) contains the dynamical
variables \(q\). Both have no effect on the Lax equation and  can thus be
omitted. They are
\begin{eqnarray*}
   A_m&=&{i\over2}(g_{L}N^m_L+g_{S}N^m_S),
   \quad \mbox{for untwisted},\qquad
   {i\over2}(g_{L}N^m_L+2g_{S}N^m_S),\quad \mbox{for twisted},
   \\
   B_m&=&{i\over2}(g_{L}N^m_LD_L+g_{S}N^m_SD_S)-
   {i\over2}\sum_{\rho\in\Delta}
   g_{|\rho|}V_{|\rho|}(\rho\cdot q),
\end{eqnarray*}
in which $N^m_{L}$ ($N^m_{S}$) is  the number of long (short) roots
$\alpha\in\Delta$ such that $\alpha\cdot\mu=0$ for a given
$\mu\in\Sigma_{min}$.
The integer
$N^m_{L}$ ($N^m_{S}$) is  well-defined since all vectors labelling the
basis elements in $\mathbf{V}$ for  the representation are in a single Weyl
orbit and hence any other vector, say
$\nu=w(\mu)$ for $w$ an element of the Weyl group.  This implies that
$\alpha\cdot\nu=0$ if and only if $w(\alpha)\cdot\mu=0$ and so $N^m_{L}$ 
is the same for any choice of $\mu\in\Sigma_{min}$. When the \(A_m\) and
\(B_m\) terms are dropped, the above Lax pair (\ref{lmin}),
(\ref{mmin}) has the same form as that given in our previous paper
\cite{Our_CM_Papers}.

For the root type
Lax pair we obtain
\begin{eqnarray}
   L_{\alpha\beta}&=&p\cdot\alpha\,\delta_{\alpha,\beta}
   +i\sum_{\rho\in\Delta}g_{|\rho|}\left[x_{|\rho|}(\rho\cdot 
   q,\xi)\delta_{\alpha-\beta,\rho}+2\,x_{|\rho|}(\rho\cdot
   q,2\xi)\delta_{\alpha-\beta,2\rho}\right]+A_r\delta_{\alpha,\beta}, 
   \label{lroot}\\
   M_{\alpha\beta}&=&i\left[g_{|\alpha|}V_{|\alpha|}(\alpha\cdot
   q)+\sum_{\Delta\ni\rho=\alpha-\sigma,\sigma\in\Delta}g_{|\rho|}
   V_{|\rho|}(\rho\cdot q)\right]\delta_{\alpha,\beta} \\ \nonumber
   &&\qquad +i\sum_{\rho\in\Delta}g_{|\rho|}\left[y_{|\rho|}(\rho\cdot
   q,\xi)\delta_{\alpha-\beta,\rho}+y_{|\rho|}(\rho\cdot
   q,2\xi)\delta_{\alpha-\beta,2\rho}\right]+B_r\delta_{\alpha,\beta}.
   \label{mroot}
\end{eqnarray}
As in the minimal case \(A_r\) is a constant and \(B_r\) contains the
dynamical variables \(q\). Both have no effect on the Lax equation and
 can be omitted. They are
\begin{eqnarray*}
   A_r&=&{i\over2}(g_{L}N^r_L+g_{S}N^r_S),
   \quad \mbox{for untwisted},\qquad
   {i\over2}(g_{L}N^r_L+2g_{S}N^r_S),\quad \mbox{for twisted},
   \\  
   B_r&=&{i\over2}(g_{L}N^r_LD_L+g_{S}N^r_SD_S)-
   {i\over2}\sum_{\rho\in\Delta}
   g_{|\rho|}V_{|\rho|}(\rho\cdot q),
\end{eqnarray*}
in which $N^r_{L}$ ($N^r_{S}$) is  the number of long (short) roots
$\alpha\in\Delta$ such that $\alpha\cdot\beta=0$ for a given
$\beta\in\Delta,(\Delta_L,\Delta_S)$. When the \(A_r\) and
\(B_r\) terms are dropped and with the following identification:
\begin{eqnarray}
   x_L(u,\xi)&\equiv& x(u,\xi),\qquad\ x_S(u,\xi)\equiv  x(u,\xi)\qquad
   \mbox{or} \quad x^{(1/2)}(u,\xi),
   \nonumber\\
   x_L(u,2\xi)&\equiv& x_d(u,\xi),
   \qquad x_S(u,2\xi)\equiv x_d(u,\xi)\quad
   \mbox{or} \quad x_d^{(1/2)}(u,\xi),
   \label{funciden}
\end{eqnarray}
the above Lax pair (\ref{lroot}), (\ref{mroot}) has the same form as that
given in our previous paper
\cite{Our_CM_Papers}.

The restrictions on the eigenvalues of the operator
(\ref{MinRepCondition}), (\ref{rootRepCondition}), simplify the
functional equations, too. Let us examine the $A_{2}$ functional equation
(\ref{A2SumRule}) by adopting the variables $u$, $v$, $\xi_{1}$, 
and $\xi_{2}$ defined in (\ref{IdentityVarDef}). We impose a condition
\(\xi_1=\xi=-\xi_2\), so that the restriction of the minimal
representation (\ref{MinRepCondition}) is satisfied.
The limit formulas (\ref{x_w_Limit}) and (\ref{y_w_Limit}) yield  the so
called first sum rule of  Ref. \cite{Our_CM_Papers}
\begin{equation}
   \label{FirstSumRule}
   x(u,\xi)\,y(v,\xi)-y(u,\xi)\,x(v,\xi)+x(u+v,\xi)[V(v)-V(u)]=0.
\end{equation} 
Here the suffix \(|\rho|\) is omitted since all the roots participating
in the $A_{2}$ functional equation belong to the same orbit.
Likewise, let us impose a condition
\(\xi_1=2\xi\), \(\xi_2=-\xi\), so that the restriction of the root type
representation (\ref{rootRepCondition}) is satisfied.
The limit formulas (\ref{x_w_Limit}) and (\ref{y_w_Limit}) yield  the so
called second sum rule of  Ref. \cite{Our_CM_Papers}
\begin{eqnarray}
   0&=&x(u+v,\xi)y(u,\xi)-y(u+v,\xi)x(u,\xi)\nonumber\\
   &&\quad +2\left[x(u,2\xi)y(v,\xi)-y(-v,\xi)x(u+v,2\xi)\right]
   \nonumber\\
   &&\qquad +x(-v,\xi)y(u+v,2\xi)-y(u,2\xi)x(v,\xi).
   \label{secsum}
\end{eqnarray}

One may also verify that all of the functional equations 
in \cite{Our_CM_Papers} which must be satisfied by the functions 
$x_{|\rho|}(u,w)$ appear as various restrictions of the operator
equation in 
(\ref{SimpleSumRule}), although we will not derive them here.
Since the functional equations for the minimal 
and root type representations
are restricted,  their solution spaces are larger than that for a
generic representation.
For example, these functional equations have the same solutions as
the  general solutions
given in section \ref{consistency}, except that the exponential factor
is changed from $\exp[b\,w\,u]$ to $\exp[b(w)\,u]$, in which an arbitrary
function
\(b(w)\) need not be linear in \(w\).

\section{Comments and Discussion}
\label{comdis}
\setcounter{equation}{0}

Firstly, let us comment on the relation between our work and the paper on
elliptic Dunkl operators by Buchstaber, Felder, 
and Veselov \cite{BFV}. From the commutativity of Dunkl operators 
they derived a functional
equation Eq. (9) of Ref. \cite{BFV}, which was closely related to our
equations \(f_{A_2}=0\), \(f_{B_2}=0\), \(f_{G_2}=0\). But their
equation did not contain the spectral parameter dependence explicitly.
They obtained what would amount to our untwisted solution for the   
\(f_{A_2}=0\) functional equation. This gives a clue that the
classical and quantum integrability of the generalised Calogero-Moser
models are  closely connected.
Secondly, some remarks about the Calogero-Moser models based on \(B_r\),
\(C_r\), and \(BC_r\) root systems. The short roots of \(B_r\), the long
roots of \(C_r\), and the long and short roots of \(BC_r\), when
restricted to any two-dimensional plane, form only an \(A_1\times A_1\)
sub-root system.
This means that their short root function \(x_S(u,w)\) (and/or
\(x_L(u,w)\)) in these models are required to satisfy the \(B_2\)
functional equation only but not the \(A_2\) one. Thus the solution space
is larger than that of the other models.
This in turn allows more potential function terms (one more for \(B_r\), and
\(C_r\) and two more for \(BC_r\)) with independent coupling constant(s)
in the Hamiltonian without breaking integrability.
We call these models extended twisted models. The root type and minimal
type Lax pairs for the extended models are given in
\cite{Our_CM_Papers,Ino}. The Lax pair operators for the
extended twisted models can be constructed in a similar way as is
given in this paper. 
Thirdly, a few words about the structure of 
the functions
\(x_{|\rho|}(\rho\cdot q,(\rho^{\vee}\!\!\cdot\mu)\xi)\) and their
functional equations (\ref{TwoDSumRules}), (\ref{SimpleSumRule}), 
and the self-duality of the two-dimensional crystallographic root systems.
The  coefficients in the functional equation (\ref{TwoDSumRules}), 
{\it i.e.\/} \(\rho^{\vee}\!\!\cdot\mu\) and
\(s_{\rho}(\sigma)^{\vee}\!\!\cdot\mu\), etc., 
come from the second argument
of the function \(x_{|\rho|}(\rho\cdot q,(\rho^{\vee}\!\!\cdot\mu)\xi)\).
Namely, they are co-roots. On the other hand, the gradient operator,
\(\partial/\partial q\), in (\ref{SimpleSumRule}) supplies the coefficient
from the first argument of the function \(x_{|\rho|}(\rho\cdot
q,(\rho^{\vee}\!\!\cdot\mu)\xi)\), that is the roots. The operator
\(R_{\pm \pi/4}\) in the \(B_2\) case and \(R_{\pm \pi/6}\) in the \(G_2\)
case rotates a short root into a long root position and vice versa.
In other words, the rotation operator \(R_{\psi/2}\) in  
(\ref{SimpleSumRule}) performs the
necessary conversion from the roots to the co-roots. 
This is possible because of the
well-known self-duality of the two-dimensional 
crystallographic root systems,
\(A_2\), \(B_2\), and \(G_2\), under 
\(\alpha\leftrightarrow \alpha^{\vee}=
{2\alpha/\alpha^2}\).
As a final remark, let us comment on the integrability of the
generalised Calogero-Moser models.
As is well known, the existence of the independent involutive
conserved quantities as many as the degrees of freedom is necessary and
sufficient for the Liouville integrability.
To the best of our knowledge, the involution of the conserved quantities
for the models based on the exceptional root systems and the
non-crystallographic root systems as well as all the twisted models
is yet to be demonstrated.

\section*{Acknowledgements}
\setcounter{equation}{0}
We thank P. Sutcliffe for bringing \cite{BFV} to our attention.
We thank S.\,P.\, Khastgir, K.\, Takasaki and A.\,P.\,Veselov for 
useful discussions.
This work is partially supported  by the Grant-in-aid from the
Ministry of Education, Science and Culture, priority area
(\#707)  ``Supersymmetry and unified theory of elementary
particles". A.\,J.\,B. is supported by the Japan Society for the
Promotion of Science  and the National Science Foundation under
grant no. 9703595.
E.\,C. thanks the UK Engineering and Physical Sciences Research Council
for travel support under grant number GR/K 79437.

\appendix
\section{Appendix: Solutions to functional equations}
\setcounter{equation}{0}

\renewcommand{\theequation}{A.\arabic{equation}}

In this appendix we demonstrate that the solutions
(\ref{Genl_A2_Soln}), (\ref{UntwistedB2Solns}), (\ref{TwistedB2Solns}),
 (\ref{TwistedG2Solns}),
(\ref{OddDihedralSoln}), (\ref{EvenDihedralSoln}) given in section
\ref{consistency} satisfy the functional
equations  for the consistency of the Lax pair
\begin{equation}
   \label{SimpleSumRuleagain}
   0 = \mu\cdot R_{\psi/2}\ {\partial\over {\partial q}}f(q,\mu).
\end{equation}
We do not give a proof that these solutions are the most general ones.
First let us remark on a symmetry of the solutions of 
(\ref{SimpleSumRuleagain}) such that 
if $x_{|\rho|}(u,w)$ is a
solution then 
\begin{equation} 
   \label{Trans_Soln}
  x_{|\rho|}^{new}(u,w)
\equiv x_{|\rho|}(u,w)\exp[b\,u\,w],\quad \forall b\in {\mathbf C},
\end{equation}
 is also a solution.  Therefore, it is necessary to prove
(\ref{fG_vanishes}) only for one representative  function
$x_{|\rho|}(u,w)$ among those related by the symmetry (\ref{Trans_Soln}).
 Assume that the function 
 $x_{|\rho|}(u,w)$ satisfies (\ref{SimpleSumRuleagain}) with $f(q,\mu)$
defined by (\ref{GenlfDef}).
The corresponding expression in (\ref{GenlfDef}) for the new solution 
$x_{|\rho|}^{new}(u,w)\equiv x_{|\rho|}(u,w)\exp[b\,u\,w]$
will be  
denoted as $f(q,\mu)^{new}$.
Then
\begin{eqnarray}
   \label{Transformedf}
   f(q,\mu)^{new}&=&\sum_{\rho\neq\sigma\in\Phi_{+},\,
   R_{\psi}=s_{\rho}s_{\sigma}}\!\!g_{|\rho|}
   \,g_{|\sigma|}\,\biggl\{x_{|\rho|}(\rho\cdot 
   q,(\rho^{\vee}\!\!\cdot\mu)\xi)\,x_{|\sigma|}(\sigma\cdot
   q,(s_{\rho}(\sigma)^{\vee}\!\!\cdot\mu)\xi) \nonumber\\ 
   &&\hspace{3cm}\times\exp\left[b\,\xi\left((\rho\cdot
   q)(\rho^{\vee}\!\!\cdot\mu)+(\sigma\cdot q)
   (s_{\rho}(\sigma)^{\vee}\!\!\cdot\mu)\right)\right]\biggr\}.
\end{eqnarray}
The exponent is proportional to
\[
q\cdot\mu-q\cdot s_{\sigma}\left(s_{\rho}(\mu)\right)\equiv
q\cdot\mu -q\cdot R_{-\psi}\mu.
\]
Two pairs of roots $(\rho,\sigma)$ and $(\rho',\sigma')$ which appear
in different terms of (\ref{Transformedf}) are related by 
$s_{\rho}s_{\sigma}=s_{\rho'}s_{\sigma'}$, which implies also
\begin{equation}
   s_{\sigma}s_{\rho}=s_{\sigma'}s_{\rho'}=R_{-\psi}.
\end{equation}
Therefore the exponential factor in (\ref{Transformedf}) is common to
all terms and may be factored out of the sum.  The functional equation
(\ref{SimpleSumRuleagain}) for $f(q,\mu)^{new}$ then is
\begin{eqnarray}
   &&\mu\cdot R_{\psi/2}\ {\partial\over{\partial q}}f(q,\mu)^{new}
   \\ \nonumber
   &=&\mu\cdot R_{\psi/2}\ {\partial\over{\partial q}}\left(f(q,\mu)
   \exp\left[b\xi (q\cdot\mu -q\cdot R_{-\psi}\mu)\right]\right) 
   \\ \nonumber
   &=&b\xi f(q,\mu)\exp\left[b\xi(q\cdot\mu -q\cdot R_{-\psi}\mu)\right]
   \left(\mu\cdot
   R_{\psi/2}\mu -\mu\cdot R_{-\psi/2}\mu\right) 
   \\ \nonumber
   &=& 0.
\end{eqnarray}
Here the orthogonality of rotation operators
\begin{equation}
   \mu\cdot
   R_{\psi/2}\mu -\mu\cdot R_{-\psi/2}\mu=0,\quad
   R_{-\psi/2}=R_{\psi/2}^T
\end{equation}
is used.
This then demonstrates that the
transformed function $x_{|\rho|}(u,w)^{new}$ satisfies the functional 
equation
if the original function $x_{|\rho|}(u,w)$ does.

Next we show in turn that all of the solutions given in section
\ref{consistency} satisfy
\begin{equation}
   \label{fG_vanishes}
   f_{G}(q,\mu) =0,\qquad G=A_2,B_2,G_2,
\end{equation}
which is a sufficient condition for the solutions of
(\ref{SimpleSumRuleagain}). Note that this is a necessary and sufficient
condition for the commutativity of the corresponding Dunkl operators
\cite{BFV}.

\subsection{$A_{2}$ solution (\ref{Genl_A2_Soln})} 
Defining the variables
\begin{eqnarray}
   \label{IdentityVarDef}
   u&=&\alpha\cdot q,\qquad\quad v = \beta\cdot q, \\ \nonumber
   \xi_{1}&=& \alpha^{\vee}\!\cdot\mu,\qquad \xi_{2} =
\beta^{\vee}\!\cdot\mu,
\end{eqnarray} 
$f_{A_{2}}(q,\mu)$ becomes
\begin{equation}
   \label{A2fSimple}
   f_{A_{2}}(q,\mu)=
   x(u+v,\xi_{1}+\xi_{2})\,x(u,-\xi_{2})+x(v,\xi_{2})
   \,x(u+v,\xi_{1})-x(u,\xi_{1})\,x(v,\xi_{1}+\xi_{2}).
\end{equation}
By substituting  the solution 
\begin{equation}
   \label{A2_Sub_Soln}
   x(u,w) = {\sigma(w-u)
   \over{\sigma(w)
   \sigma(u)}}
\end{equation}
it reads, after extracting a common denominator, 
\begin{eqnarray}
   f_{A_{2}}(u,v,\xi_{1},\xi_{2})&=&
   \bigl[\sigma(\xi_{1}+\xi_{2}-u-v)\,\sigma(\xi_{2}+u)\,\sigma(v)
   \,\sigma(\xi_{1}) \\ \nonumber
   &&+\sigma(\xi_{2}-v)\,\sigma(\xi_{1}-u-v)\,\sigma(u)
   \,\sigma(\xi_{1}+\xi_{2}) \\ \nonumber
   &&-\sigma(\xi_{1}-u)\,\sigma(\xi_{1}+\xi_{2}-v)\,\sigma(u+v)
   \,\sigma(\xi_{2})\bigr] \\ \nonumber
   &&\ \ /\left[\sigma(u)\,\sigma(v)\,\sigma(u+v)\,\sigma(\xi_{1})
   \sigma(\xi_{2})\,\sigma(\xi_{1}+\xi_{2})\right].
\end{eqnarray}
The expression in the numerator may be shown to vanish, 
by using the following identity (see page 153 of \cite{Lawden} and Eq.
(5.1) of 
\cite{BrCal})
\begin{eqnarray}
   \label{Sigma_Fn_Identity}
   0 &=& \,\,\,\,\sigma(z-u_{1})\,\sigma(z+u_{1})\,\sigma(u_{2}-u_{3})
   \,\sigma(u_{2}+u_{3}) \\ \nonumber
   &&+ \sigma(z-u_{2})\,\sigma(z+u_{2})\,\sigma(u_{3}-u_{1})
   \,\sigma(u_{3}+u_{1}) \\ \nonumber
   &&+ \sigma(z-u_{3})\,\sigma(z+u_{3})\,\sigma(u_{1}-u_{2})
   \,\sigma(u_{1}+u_{2}) 
\end{eqnarray}
with the choice of variables
\begin{eqnarray}
   z &=& \xi_{2}+{1\over 2}(\xi_{1}-v), 
   \qquad u_{1}=u-{1\over 2}(\xi_{1}-v), \\
   \nonumber
   u_{2} &=& {1\over 2}(\xi_{1}+v), 
   \qquad\qquad u_{3}={1\over 2}(\xi_{1}-v).
\end{eqnarray}
This implies that $f_{A_{2}}(q,\mu)=0$ for the function in
(\ref{A2_Sub_Soln}) and after including the symmetry of 
solutions (\ref{Trans_Soln}), the 
more general function in (\ref{Genl_A2_Soln}) also gives 
$f_{A_{2}}(q,\mu)=0$ and therefore the $A_{2}$ functional equation is
satisfied.

\subsection{$B_{2}$ untwisted solution (\ref{UntwistedB2Solns})} 
Next we turn to the functional equation arising from the \(B_2\) 
sub-root system. This, as well as the functional equation associated with
the 
\(G_2\) root system, admits two types of solutions, the untwisted and the
twisted ones. By using the same definitions of the variables $u$, $v$,
$\xi_{1}$,  and $\xi_{2}$ as in (\ref{IdentityVarDef}), except for the
fact that $\alpha$ and $\beta$ are the short and long
simple roots of $B_{2}$, respectively, the expression for
$f_{B_{2}}(q,\mu)$ becomes
\begin{eqnarray}
   \label{B2fSimple}
   f_{B_{2}}(u,v,\xi_{1},\xi_{2}) =
   &-&x_{S}(u,\xi_{1})\,x_{L}(v,\xi_{1}+\xi_{2}) \\ \nonumber
   &+&x_{S}(u+v,\xi_{1}+2\xi_{2})\,x_{L}(2u+v,-\xi_{2}) \\ \nonumber
   &+&x_{S}(u,-\xi_{1}-2\xi_{2})\,x_{L}(2u+v,\xi_{1}+\xi_{2}) 
   \\ \nonumber
   &+&x_{S}(u+v,\xi_{1})\,x_{L}(v,\xi_{2}).
\end{eqnarray}
Substituting in the untwisted solutions 
\begin{equation}
   \label{B2_Untwisted_Sub_Soln}
   x_{L}(u,w)=x_{S}(u,w)= {\sigma(w-u)
   \over{\sigma(w)
   \sigma(u)}}
\end{equation}
this becomes
\begin{eqnarray}
   f_{B_{2}}(u,v,\xi_{1},\xi_{2}) &=& 
   \bigl[-\sigma(\xi_{1}-u)\,\sigma(\xi_{1}+\xi_{2}-v)\,\sigma(u+v)
   \,\sigma(2u+v)\,\sigma(\xi_{2})\,\sigma(\xi_{1}+2\xi_{2}) 
   \nonumber\\ 
   &&+
   \sigma(\xi_{1}+2\xi_{2}-u-v)\,\sigma(\xi_{2}+2u+v)\,\sigma(u)
   \,\sigma(v)\,\sigma(\xi_{1})\,\sigma(\xi_{1}+\xi_{2}) 
   \nonumber\\ \nonumber
   &&+\sigma(\xi_{1}+2\xi_{2}+u)\,\sigma(\xi_{1}+\xi_{2}-2u-v)
   \,\sigma(v)\,\sigma(u+v)\,\sigma(\xi_{1})\,\sigma(\xi_{2}) 
   \\ \nonumber
   &&+\sigma(\xi_{1}-u-v)\,\sigma(\xi_{2}-v)\,\sigma(u)\,\sigma(2u+v)
   \,\sigma(\xi_{1}+\xi_{2})\,\sigma(\xi_{1}+2\xi_{2})\bigr] \\ 
   &&\ \ /\left[\sigma(u)\,\sigma(v)
   \sigma(2u+v)\,\sigma(\xi_{1})\,\sigma(\xi_{2})\,
   \sigma(\xi_{1}+\xi_{2})\,
   \sigma(\xi_{1}+2\xi_{2})\right].
\end{eqnarray}
Let us denote the
numerator by $g_{B_{2}}(u,v,\xi_{1},\xi_{2})$.  Gathering terms, one
obtains
\begin{eqnarray}
   g_{B_{2}}(u,v,\xi_{1},\xi_{2})&=&\sigma(2u+v)\,
   \sigma(\xi_{1}+2\xi_{2})\bigl[-\sigma(\xi_{1}-u)
   \,\sigma(\xi_{1}+\xi_{2}-v)\,\sigma(u+v)\,\sigma(\xi_{2}) 
   \nonumber\\ \nonumber
   &&+\sigma(\xi_{1}-u-v)\,\sigma(\xi_{2}-v)\,\sigma(u)
   \,\sigma(\xi_{1}+\xi_{2})\bigr] \\ \nonumber
   &&+\sigma(v)\,\sigma(\xi_{1})\bigl[\sigma(\xi_{1}+2\xi_{2}-u-v)
   \,\sigma(\xi_{2}+2u+v)\,\sigma(u)\,\sigma(\xi_{1}+\xi_{2}) 
   \\ 
   &&+\sigma(\xi_{1}+2\xi_{2}+u)\,\sigma(\xi_{1}+\xi_{2}-2u-v)
   \,\sigma(u+v)\,\sigma(\xi_{2})\bigr].
\end{eqnarray}
Using the identity (\ref{Sigma_Fn_Identity}) for the expression 
in the first 
set of brackets with the choice of variables
\begin{eqnarray}
   z&=&-v+{1\over 2}(-u+\xi_{1}+\xi_{2}), \qquad
u_{1}={1\over 2}(u-\xi_{1}+\xi_{2}),
   \\ \nonumber
   u_{2}&=&{1\over 2}(u+\xi_{1}+\xi_{2}), \qquad\qquad\quad\ 
   u_{3}={1\over 2}(-u+\xi_{1}+\xi_{2}),
\end{eqnarray}
and for the expression in the second set of brackets  with
the variables
\begin{eqnarray}
   z&=&{1\over 2}(u+\xi_{1}+3\xi_{2}),
   \qquad u_{1}={1\over 2}(3u+2v-\xi_{1}-\xi_{2}),
   \\ \nonumber
   u_{2}&=&{1\over 2}(u+\xi_{1}+\xi_{2}), 
   \qquad\ u_{3}={1\over 2}(-u+\xi_{1}+\xi_{2}),
\end{eqnarray}
gives
\begin{eqnarray}
   g_{B_{2}}(u,v,\xi_{1},\xi_{2})
   &=&-\sigma(2u+v)\,\sigma(\xi_{1}+2\xi_{2})\,\sigma(v)\,
   \sigma(\xi_{1}+\xi_{2}-u-v)\,\sigma(\xi_{1})\,\sigma(\xi_{2}+u) 
    \nonumber\\
   &&+\sigma(v)\,\sigma(\xi_{1})\,\sigma(\xi_{2}+u)\,
   \sigma(\xi_{1}+2\xi_{2})\,\sigma(\xi_{1}+\xi_{2}-u-v)\,\sigma(2u+v) 
   \nonumber\\ 
   &=& 0.
\end{eqnarray}
Including the possible symmetry transformations (\ref{Trans_Soln}) 
of the function (\ref{B2_Untwisted_Sub_Soln}) 
this then proves that $f_{B_{2}}(q,\mu)=0$ for the general elliptic
untwisted solution of (\ref{UntwistedB2Solns}).

The untwisted solution (\ref{UntwistedB2Solns}) of the $G_{2}$
functional equation
(\ref{G2SumRule}) may be proven again using only the $\sigma$ function
identity (\ref{Sigma_Fn_Identity}).  Since the method of proof is
essentially the same
as for the $B_{2}$ functional equation we omit the details of the proof.

\subsection{$B_{2}$ twisted solution (\ref{TwistedB2Solns})}
We next demonstrate that the twisted solution in (\ref{TwistedB2Solns})
satisfies the $B_{2}$ functional equation (\ref{B2SumRule}).  
First we define a particular untwisted solution $x_{0}(u,w)$
of (\ref{UntwistedB2Solns})
by assigning a special  value of the constant $b=\eta_{1}/\omega_{1}$ 
\begin{equation}
   x_{S}(u,w) = x_{L}(u,w)=x_{0}(u,w)\equiv {\sigma(w-u)
   \over{\sigma(w)
   \sigma(u)}}
   \exp[(\eta_{1}/\omega_{1})\,w\,u],
\end{equation}
where $\eta_{1}$ is defined in terms of the Weierstrass $\zeta$
function as $\eta_{1}\equiv
\zeta(\omega_{1})$.
The value of $b$ is chosen so that $x_{0}(u,w)$ is periodic in the 
$\omega_{1}$ direction
\begin{equation}
   \label{x0Periodicity}
   x_{0}(u+2\omega_{1},w)=x_{0}(u,w).
\end{equation}
Adding a constant vector to $q$ does not affect the
equation $f_{B_{2}}(q,\mu)= 0$ so $q$ is shifted as   
\begin{equation}
   q\rightarrow q+{2\omega_1\Lambda^{(\alpha)}\over |\alpha|^{2}},
\end{equation}
in which $\Lambda^{(\alpha)}$ is the fundamental weight dual to
the simple short root
$\alpha$, {\it i.e.} 
\begin{equation}
   \alpha^{\vee}\!\cdot \Lambda^{(\alpha)} = 1, \qquad 
   \beta^{\vee}\!\cdot \Lambda^{(\alpha)} = 0.
\end{equation}
Then the equation $f_{B_{2}}(q,\mu)=0$ reads
\begin{eqnarray}
   0 =&-& x_{0}(\alpha\cdot
   q+\omega_{1},\alpha^{\vee}\!\cdot\mu)\,
   x_{0}(\beta\cdot q,(2\alpha+\beta)^{\vee}\!\cdot\mu)
   \nonumber\\ \nonumber
   &+&x_{0}((\alpha+\beta)\cdot 
   q+\omega_{1},(\alpha+\beta)^{\vee}\!\cdot\mu)\,
   x_{0}((2\alpha+\beta)\cdot
   q,-\beta^{\vee}\!\cdot\mu) \\ \nonumber
   &+&x_{0}(\alpha\cdot  
   q+\omega_{1},-(\alpha+\beta)^{\vee}\!\cdot\mu)\,
   x_{0}((2\alpha+\beta)\cdot
   q,(2\alpha+\beta)^{\vee}\!\cdot\mu) \\ 
   &+&x_{0}((\alpha+\beta)\cdot
   q+\omega_{1},\alpha^{\vee}\!\cdot\mu)\,x_{0}(\beta\cdot
   q,\beta^{\vee}\!\cdot\mu),
\end{eqnarray}
where the periodicity of $x_{0}(u,w)$, (\ref{x0Periodicity}), is used.
This simply means
that 
\begin{eqnarray}
   x_{L}(u,w) &=& {\sigma(w-u)
   \over{\sigma(w)
   \sigma(u)}}
   \exp[b\,w\,u], \nonumber\\ 
   x_{S}(u,w)&=&{\sigma(w-u-\omega_{1})
   \over{\sigma(w)
   \sigma(u+\omega_{1})}}
   \exp[\eta_{1}\,w+b\,w\,u],
   \label{ShiftedB2Solns}
\end{eqnarray}
satisfy the equation $f_{B_{2}}(q,\mu)= 0$.  Since $x_{L}(u,w)$
here 
is the same function as in (\ref{UntwistedB2Solns}), 
any linear combinations of
$x_{S}(u,w)$ from the untwisted solution in (\ref{UntwistedB2Solns})
and in (\ref{ShiftedB2Solns}) also satisfy the $B_{2}$ functional
equation.  Requiring that the linear combination should satisfy
the \(A_2\) functional equation and that it has a simple pole
with unit residue at
$u=0$, we obtain
\begin{equation}
   x_{S}^{0}(u,w)=\biggl[{\sigma(w-u)
   \over{\sigma(w)
   \sigma(u)}}
   +{\sigma(w-u-\omega_{1})
   \over{\sigma(w)
   \sigma(u+\omega_{1})}}
   \exp[\eta_{1}\,w]\biggr]\exp[b\,w\,u].
\end{equation}
This expression for $x_{S}^{0}(u,w)$ has the following monodromies in $u$
and $w$:
\begin{eqnarray}
   x_{S}^{0}(u+\omega_{1},w) &=&
   x_{S}^{0}(u,w)\exp\left[(-\eta_{1}+b\,\omega_{1})w\right], 
  \nonumber\\
   \nonumber x_{S}^{0}(u+2\omega_{3},w) &=&
   x_{S}^{0}(u,w)\exp\left[2(-\eta_{3}+b\,\omega_{3})w\right], 
   \\ \nonumber
   x_{S}^{0}(u,w+2\omega_{1}) &=&
   x_{S}^{0}(u,w)\exp\left[2(-\eta_{1}+b\,\omega_{1})u\right], \\ 
   x_{S}^{0}(u,w+4\omega_{3}) &=&
   x_{S}^{0}(u,w)\exp\left[4(-\eta_{3}+b\,\omega_{3})u\right].
\end{eqnarray}
It also has the following poles and zeros in the fundamental regions
of $u$ and $w$: simple
poles at $u=0$ and $w=0$ with residues $1$ and $-2$, respectively, and
a zero at $u=w/2$.  
It may also be shown that the twisted solution to
the $B_{2}$ functional equation
$x_{S}(u,w)$ in (\ref{TwistedB2Solns}) has the same monodromies and
poles.  This implies that the ratio $x_{S}^{0}(u,w)/x_{S}(u,w)$ is an
elliptic function in both $u$ and $w$ and has no poles and therefore
is a constant.  Since the residues are the same at all the poles, the
ratio is equal to one and $x_{S}^{0}(u,w)$ is, in fact, the same as
the twisted solution in (\ref{TwistedB2Solns})

\subsection{$G_{2}$ twisted solution (\ref{TwistedG2Solns})}
The proof that the twisted solutions (\ref{TwistedG2Solns}) to 
the $G_{2}$ functional equation (\ref{G2SumRule}) follow from the
untwisted solutions proceeds in a similar manner.  
We start from  the particular untwisted solution 
$x_{S}(u,w)=x_{L}(u,w)=x_{0}(u,w)$ satisfying $f_{G_{2}}(q,\mu)= 0$,
 with the periodicity (\ref{x0Periodicity}),
and  shift $q$ as
\begin{equation}
   q\rightarrow q+{4\omega_{1}\Lambda^{(\alpha)}\over{3|\alpha|^{2}}},
\end{equation}
in which $\Lambda^{(\alpha)}$ is the fundamental weight dual to
the simple short root
$\alpha$. We  obtain
\begin{eqnarray}
   0&=&\ \ x_{0}(\alpha\cdot
   q+{2\omega_{1}\over 3},\alpha^{\vee}\!\cdot\mu)\,
   x_{0}(\beta\cdot q,(3\alpha+\beta)^{\vee}\!\cdot\mu)
   \nonumber\\ \nonumber
   && + x_{0}(\alpha\cdot
   q+{2\omega_{1}\over 3},-(2\alpha+\beta)^{\vee}\!\cdot\mu)\,
   x_{0}((3\alpha+\beta)\cdot
   q,(3\alpha+\beta)^{\vee}\!\cdot\mu) \\ \nonumber
   &&+ x_{0}((2\alpha+\beta)\cdot
   q+{4\omega_{1}\over 3},(2\alpha+\beta)^{\vee}\!\cdot\mu)\,
   x_{0}((3\alpha+\beta)\cdot
   q,-(3\alpha+2\beta)^{\vee}\!\cdot\mu) \\ \nonumber
   &&+ x_{0}((2\alpha+\beta)\cdot
   q+{4\omega_{1}\over 3},-(\alpha+\beta)^{\vee}\!\cdot\mu)\,
   x_{0}((3\alpha+2\beta)\cdot
   q,(3\alpha+2\beta)^{\vee}\!\cdot\mu) \\ \nonumber
   &&+ x_{0}((\alpha+\beta)\cdot
   q+{2\omega_{1}\over 3},(\alpha+\beta)^{\vee}\!\cdot\mu)\,
   x_{0}((3\alpha+2\beta)\cdot
   q,-\beta^{\vee}\!\cdot\mu)\\ 
   &&+ x_{0}((\alpha+\beta)\cdot 
   q+{2\omega_{1}\over 3},\alpha^{\vee}\!\cdot\mu)\,
   x_{0}(\beta\cdot q,\beta^{\vee}\!\cdot\mu).
   \label{ShiftedG2Solns1}
\end{eqnarray} 
Shifting $q$ again by the same amount, or in one step
\begin{equation}
   q\rightarrow q+{8\omega_{1}\Lambda^{(\alpha)}\over{3|\alpha|^{2}}},
\end{equation}
 gives the following equation
\begin{eqnarray}
   0&=&\ \ x_{0}(\alpha\cdot
   q+{4\omega_{1}\over 3},\alpha^{\vee}\!\cdot\mu)\,
   x_{0}(\beta\cdot q,(3\alpha+\beta)^{\vee}\!\cdot\mu)
   \nonumber\\ \nonumber
   &&+ x_{0}(\alpha\cdot
   q+{4\omega_{1}\over 3},-(2\alpha+\beta)^{\vee}\!\cdot\mu)\,
   x_{0}((3\alpha+\beta)\cdot
   q,(3\alpha+\beta)^{\vee}\!\cdot\mu) \\ \nonumber
   &&+ x_{0}((2\alpha+\beta)\cdot
   q+{2\omega_{1}\over 3},(2\alpha+\beta)^{\vee}\!\cdot\mu)\,
   x_{0}((3\alpha+\beta)\cdot
   q,-(3\alpha+2\beta)^{\vee}\!\cdot\mu) \\ \nonumber
   &&+ x_{0}((2\alpha+\beta)\cdot
   q+{2\omega_{1}\over 3},-(\alpha+\beta)^{\vee}\!\cdot\mu)\,
   x_{0}((3\alpha+2\beta)\cdot
   q,(3\alpha+2\beta)^{\vee}\!\cdot\mu) \\ \nonumber
   &&+ x_{0}((\alpha+\beta)\cdot
   q+{4\omega_{1}\over 3},(\alpha+\beta)^{\vee}\!\cdot\mu)\,
   x_{0}((3\alpha+2\beta)\cdot
   q,-\beta^{\vee}\!\cdot\mu)\\ 
   &&+ x_{0}((\alpha+\beta)\cdot 
   q+{4\omega_{1}\over 3},\alpha^{\vee}\!\cdot\mu)\,
   x_{0}(\beta\cdot q,\beta^{\vee}\!\cdot\mu),
   \label{ShiftedG2Solns2}
\end{eqnarray}
where the periodicity of $x_{0}(u,w)$, (\ref{x0Periodicity}), is also
used.  Adding  (\ref{ShiftedG2Solns1}) and
(\ref{ShiftedG2Solns2}) together and using the symmetry of solutions in 
(\ref{Trans_Soln}) implies that the following functions
are solutions to the $G_{2}$ functional equation (\ref{G2SumRule}):
\begin{eqnarray}
   \label{ShiftedG2Solns}
   x_{L}(u,w) &=& {\sigma(w-u)
   \over{\sigma(w)
   \sigma(u)}}
   \exp[b\,w\,u], \\ \nonumber
   x_{S}(u,w) &=& \biggl[{\sigma(w-u-{2\omega_{1}\over 3})
   \over{\sigma(w)
   \sigma(u+{2\omega_{1}\over
   3})}}\exp[(2/3)\eta_{1}\,w] 
   + {\sigma(w-u-{4\omega_{1}\over 3})
   \over{\sigma(w)
   \sigma(u+{4\omega_{1}\over
   3})}}\exp[(4/3)\eta_{1}\,w]\biggr]
   \exp[b\,w\,u].
\end{eqnarray}
Again, since $x_{L}(u,w)$ is the same function as in
(\ref{UntwistedB2Solns}) an arbitrary linear combination of the
$x_{S}(u,w)$ from (\ref{UntwistedB2Solns}) and (\ref{ShiftedG2Solns})
also satisfies the $G_{2}$ functional equation.  
Requiring that the linear combinations should satisfy
the \(A_2\) functional equation and that it has a simple pole
with unit residue at
$u=0$, we obtain
\begin{eqnarray}
   x_{S}^{0}(u,w) &=& \biggl[{\sigma(w-u)
   \over{\sigma(w)
   \sigma(u)}}
   +{\sigma(w-u-{2\omega_{1}\over 3})
   \over{\sigma(w)
   \sigma(u+{2\omega_{1}\over
   3})}}\exp[(2/3)\eta_{1}\,w] \\
   \nonumber
   &&\hspace{2cm} + {\sigma(w-u-{4\omega_{1}\over 3}
   )
   \over{\sigma(w)
   \sigma(u+{4\omega_{1}\over
   3})}}\exp[(4/3)\eta_{1}\,w]\biggr]
   \exp[b\,w\,u].
\end{eqnarray}
This function has the following monodromies:
\begin{eqnarray}
   x_{S}^{0}(u+{2\omega_{1}\over 3},w) &=&
   x_{S}^{0}(u,w)\exp\left[{2\over 3}(-\eta_{1}+b\,\omega_{1})w\right],
\\ 
   \nonumber
   x_{S}^{0}(u+2\omega_{3},w) &=&
   x_{S}^{0}(u,w)\exp\left[2(-\eta_{3}+b\,\omega_{3})w\right], \\
   \nonumber
   x_{S}^{0}(u,w+2\omega_{1}) &=&
   x_{S}^{0}(u,w)\exp\left[2(-\eta_{1}+b\,\omega_{1})u\right], \\
   \nonumber
   x_{S}^{0}(u,w+6\omega_{3}) &=&
   x_{S}^{0}(u,w)\exp\left[6(-\eta_{3}+b\,\omega_{3})u\right].
\end{eqnarray}
The following properties of $x_{S}^{0}(u,w)$ can be shown. It has poles
and zeros in the fundamental regions of $u$ and $w$: simple poles at $u=0$
and $w=0$ with residues $1$ and $-3$, respectively, and a zero at $u=w/3$.
Since it may be shown that the twisted solution $x_{S}(u,w)$ to the
$G_{2}$ functional equation in (\ref{TwistedG2Solns}) has the same
monodromies, poles and zeros as $x_{S}^{0}(u,w)$, by the same argument
as in the $B_{2}$ case given above, $x_{S}^{0}(u,w)=x_{S}(u,w)$ and
the solutions derived above are the same as the 
twisted solutions in (\ref{TwistedG2Solns}). 

\subsection{Dihedral solutions (\ref{OddDihedralSoln}), 
(\ref{EvenDihedralSoln})}
Next, it will be shown that the solutions (\ref{OddDihedralSoln})
and (\ref{EvenDihedralSoln}) 
satisfy the dihedral $I_{2}(m)$ functional equations 
(\ref{I2mSumRule}) for  odd or even $m$, respectively.  We assume that 
all roots are of the same length, even for an even
integer $m$, since they may be made so by a redefinition of the coupling
constants.  
First consider the case of  an odd integer $m$ and arbitrary 
$N=1,\ldots,[m/2]$.  Substituting the functions
\begin{equation}
   \label{OddDihedral_Sub_Soln}
   x(u,w)=\left({1\over u}-{1\over w}\right)
\end{equation} 
in  (\ref{I2mfDef}) for 
$f_{I_{2}(m)}^{N}(q,\mu)$ and redefining 
$\xi\rightarrow \xi |\alpha_{j}|^{2}/2$ gives
\begin{eqnarray}
   f_{I_{2}(m)}^{N}(q,\mu)&=&g^{2}\sum_{j=1}^{m}x(\alpha_{j+N}\cdot
   q,(\alpha_{j+N}\cdot \mu)\xi)\,x(\alpha_{j}\cdot
   q,-(\alpha_{j+2N}\cdot\mu)\xi) \nonumber\\ \nonumber
   &=&g^{2}\sum_{j=1}^{m}\left({1\over{\alpha_{j+N}\cdot
   q}}-{1\over{(\alpha_{j+N}\cdot\mu)\xi}}\right)
   \left({1\over{\alpha_{j}\cdot 
   q}}+{1\over{(\alpha_{j+2N}\cdot\mu)\xi}}\right) \\ \nonumber
   &=&g^{2}\sum_{j=1}^{m}\Biggl[{1\over{(\alpha_{j+N}\cdot
   q)(\alpha_{j}\cdot q)}}+{1\over{(\alpha_{j+N}\cdot
   q)(\alpha_{j+2N}\cdot\mu)\xi}} \\ 
   &&\quad\quad\quad -{1\over{(\alpha_{j}\cdot
   q)(\alpha_{j+N}\cdot\mu)\xi}}-{1\over{(\alpha_{j+N}\cdot\mu)
   (\alpha_{j+2N}\cdot\mu)\xi^{2}}}\Biggr].
\end{eqnarray}
The $O(1/\xi)$ terms cancel pairwise using the identity
\begin{equation}
   \sum_{j=1}^{m}{1\over{(\alpha_{j+d}\cdot
   q)(\alpha_{j+N+d}\cdot\mu)}}
   =\sum_{j=1}^{m}{1\over{(\alpha_{j+d+h}\cdot q)
   (\alpha_{j+d+N+h}\cdot\mu)}}
\end{equation}
with $h$, $d$, and $N$ arbitrary integers. 
This identity follows from 
the property of the $I_{2}(m)$ roots 
$\alpha_{j+m}=-\alpha_{j}$. 
The vanishing of the $O(\xi^{0})$ and
$O(1/\xi^{2})$ terms follows from 
a simple trigonometric identity (\(q=|q|(\cos t,\sin t)\)):
\begin{equation}
   \sum_{j=1}^{m}{1\over{\cos(t-{j\pi/ m})\cos(t-{(j+N)\pi/ m})}}
   =0.
   \label{trigiden}
\end{equation}
The left hand side is a meromorphic function in $t$ with a period $\pi$
and it is exponentially decreasing as $t\to\pm i\infty$.
It is elementary to show that all the residues of 
the possible simple poles
${\pi\over2}+{j\pi\over m}$, $j=1,2,\ldots,m$ vanish.
Thus (\ref{trigiden}) vanishes.
This shows that the dihedral 
functional equation (\ref{I2mSumRule}) for odd $m$  is satisfied by
the functions in (\ref{OddDihedral_Sub_Soln}).   Including the
symmetry of solutions of (\ref{Trans_Soln}) implies that the more general
solution in (\ref{OddDihedralSoln}) also satisfies (\ref{I2mSumRule}).

Next consider the case of even $m$ and odd $N$ in
the functional equation (\ref{I2mSumRule}).
Substituting the functions 
\begin{equation}
   x_{O}(u,w)=\left({1\over u}-{1\over w}\right), \qquad
   x_{E}(u,w)=\left({1\over u}-{c\over w}\right)
\end{equation}
in the 
equation for $f_{I_{2}(m)}^{N}(q,\mu)$ (\ref{I2mfDef}) 
and redefining $\xi\rightarrow \xi |\alpha_{j}|^{2}/2$, as before, gives
\begin{eqnarray}
   f_{I_{2}(m)}^{N}(q,\mu)&=& g_{O}g_{E}\sum_{j=1}^{m/2}
   \biggl[x_{E}(\alpha_{2j}\cdot
   q,-(\alpha_{2j+2N}\cdot\mu)\xi)\,x_{O}(\alpha_{2j+N}\cdot
   q,(\alpha_{2j+N}\cdot\mu)\xi) \nonumber\\ \nonumber
   &&\qquad\qquad +x_{E}(\alpha_{2j+N-1}\cdot q,(\alpha_{2j+N-1}\cdot
   \mu)\xi)\,x_{O}(\alpha_{2j-1}\cdot
   q,-(\alpha_{2j+2N-1}\cdot\mu)\xi)\biggr] \\
   &=& g_{O}g_{E}\Biggl\{\sum_{j=1}^{m}\left[{1\over{(\alpha_{j}\cdot
   q)(\alpha_{j+N}\cdot
   q)}}-{c\over{(\alpha_{j+2N}\cdot\mu)
   (\alpha_{j+N}\cdot\mu)\xi^{2}}}\right] 
   \nonumber\\ \nonumber
   &&\quad\quad\quad+{1\over
   \xi}\sum_{j=1}^{m/2}\Biggl[{c\over{(\alpha_{2j+N}\cdot 
   q)(\alpha_{2j+2N}\cdot\mu)}}-{c\over{(\alpha_{2j-1}\cdot q)
   (\alpha_{2j+N-1}\cdot\mu)}} \\
   &&\quad\quad\quad-{1\over{(\alpha_{2j}\cdot q)(\alpha_{2j+N}\cdot\mu)}}
   +{1\over{(\alpha_{2j+N-1}\cdot q)(\alpha_{2j+2N-1}\cdot\mu)}}
   \Biggr]\Biggr\}.
\end{eqnarray}
As before, the $O(1/\xi)$ terms in $f_{I_{2}(m)}^{N}(q,\mu)$ cancel
pairwise using
the identity
\begin{equation}
   \sum_{j=1}^{m/2}{1\over{(\alpha_{2j+d}\cdot
   q)(\alpha_{2j+N+d}\cdot\mu)}}
   =\sum_{j=1}^{m/2}{1\over{(\alpha_{2j+d+h}\cdot q)
   (\alpha_{2j+d+N+h}\cdot\mu)}}
\end{equation}
with $h$ an even integer and $d$ and $N$ arbitrary integers.  
The $O(\xi^{0})$ and
$O(1/\xi^{2})$ terms are proportional to the corresponding terms for
$m$ odd given above and therefore vanish.  Including the symmetry of
(\ref{Trans_Soln}), the solution of (\ref{EvenDihedralSoln})
satisfies (\ref{I2mSumRule}) for $m$ even and $N$ odd.

For the case of the solutions for the $I_{2}(m)$ functional equation 
for even $m$  
and even $N$, note that $f_{I_{2}(m)}^{N}(q,\mu)$ may be
written as
\begin{eqnarray}  
f_{I_{2}(m)}^{N}(q,\mu)&=&\sum_{j=1}^{m/2}
   \biggl[g_O^2x_{O}(\alpha_{2j+N}\cdot
   q,(\alpha_{2j+N}\cdot\mu)\xi)\,x_{O}(\alpha_{2j}\cdot
   q,-(\alpha_{2j+2N}\cdot\mu)\xi) \nonumber\\ \nonumber
   &&\quad\ \ +g_E^2x_{E}(\alpha_{2j+N-1}\cdot
   q,(\alpha_{2j+N-1}\cdot\mu)\xi)\,x_{E}(\alpha_{2j-1}\cdot
   q,-(\alpha_{2j+2N-1}\cdot\mu)\xi)\biggr] \\ \nonumber
   &=& f_{I_{2}(m/2)}^{N/2}(q,\mu)+g_{I_{2}(m/2)}^{N/2}(q,\mu) \\ 
   &=& 0.
\end{eqnarray}
Here $g_{I_{2}(m/2)}^{N/2}(q,\mu)$ is proportional to
$f_{I_{2}(m/2)}^{N/2}(q,\mu)$ with the $I_{2}(m/2)$ roots rotated
by $\pi/m$ and so the vanishing of $f_{I_{2}(m)}^{N}(q,\mu)$ follows 
using the previous equation and induction on $N$.
Therefore the solutions in (\ref{EvenDihedralSoln}) solve the $I_{2}(m)$ 
functional equation (\ref{I2mSumRule}) for all $m$ and $N$.


\end{document}